\documentclass[aps,prx,twocolumn,showpacs,superscriptaddress,floatfix]{revtex4-2}  

\usepackage[letterpaper,top=1in,bottom=1in,left=1in,right=1in,marginparwidth=1in]{geometry}

\usepackage{amsmath}
\usepackage{newtxtext,newtxmath}
\usepackage{graphicx}
\usepackage[colorlinks=true, allcolors=blue]{hyperref}
\usepackage{indentfirst}

\usepackage{tikz,tikz-3dplot}
\usetikzlibrary{patterns}
\usepackage{graphicx,float}

\usepackage{bbold}

\definecolor{darkred}{rgb}{0.7,0.0,0.0}

\definecolor{darkblue}{rgb}{0,0.02,0.45}

\usepackage{array}   
\usepackage{makecell}
\newcolumntype{C}{>{$}c<{$}} 
\newcolumntype{R}{>{$}r<{$}} 
\usepackage{ucs}
\usepackage{natbib}
\usepackage{graphicx}  
\usepackage{bm}        
\usepackage{amsmath}
\usepackage{color}
\usepackage{CJK}
\usepackage{times}
\usepackage{verbatim}
\usepackage{mathrsfs}
\usepackage{hyperref}
\usepackage{ulem}
\usepackage{physics}
\usepackage{newtxtext,newtxmath}
\hypersetup{colorlinks = True, urlcolor=blue, linkcolor=blue, citecolor=blue}

\newcommand{\bg}{\begin{pmatrix}}
\newcommand{\ed}{\end{pmatrix}}
\newcommand{\dg}{\dagger}
\newcommand{\sg}{\sigma}
\newcommand{\al}{\alpha}
\newcommand{\bt}{\beta}

\newcommand{\gm}{\gamma}
\newcommand{\ep}{\epsilon}
\newcommand{\dt}{\delta}
\newcommand{\ta}{\theta}
\newcommand{\bsym}{\boldsymbol}
\newcommand{\pr}{\prime}
\newcommand{\sumgm}{\sum_{\left\langle ij\right\rangle _{\gm}}}
\newcommand{\lelagm}{\left\langle ij\right\rangle _{\gm}}

\newcommand{\mb}[1]{\mathbf{#1}}

\usepackage{cancel}
\usepackage{comment}

\begin{document}

\title{Ferrimagnetic Kitaev spin liquids in mixed spin 1/2 spin 3/2 honeycomb magnets}

\author{Willian Natori}
\affiliation{Gleb Wataghin Institute of Physics, University of Campinas, Campinas,
São Paulo 13083-950, Brazil}
\affiliation{Institute Laue-Langevin, BP 156, 41 Avenue des Martyrs, 38042 Grenoble
Cedex 9, France}
\author{Yang Yang}
\affiliation{School of Physics and Astronomy, University of Minnesota, Minneapolis, MN 55455, USA}
\affiliation{Department of Physics, University of Virginia, Charlottesville, VA 22904, USA}
\author{Hui-Ke Jin}
\affiliation{Technical University of Munich, TUM School of Natural Sciences, Physics Department, 85748 Garching, Germany}
\affiliation{Munich Center for Quantum Science and Technology (MCQST), Schellingstr. 4, 80799 M{\"u}nchen, Germany}
\affiliation{School of Physical Science and Technology, ShanghaiTech University, Shanghai 201210, China}
\author{Johannes Knolle}
\affiliation{Technical University of Munich, TUM School of Natural Sciences, Physics Department, 85748 Garching, Germany}
\affiliation{Munich Center for Quantum Science and Technology (MCQST), Schellingstr. 4, 80799 M{\"u}nchen, Germany}
\affiliation{Blackett Laboratory, Imperial College London, London SW7 2AZ, United Kingdom}
\author{Natalia B. Perkins}
\affiliation{School of Physics and Astronomy, University of Minnesota, Minneapolis, MN 55455, USA}

\date{\today}
\begin{abstract} 
We explore the phase diagram of a mixed-spin Kitaev model, where spin-1/2 and spin-3/2 ions form a staggered pattern on a honeycomb lattice. Enabled by an exact mapping of local conserved flux operators onto $Z_2$ gauge fields, we perform a parton mean-field theory for the model with a single-ion anisotropy. The phase diagram contains four types of quantum spin liquids distinguished by quadrupolar  parameters. These analytical results are quantitatively confirmed by state-of-the-art DMRG simulations. We also explore the potential experimental realization of the mixed-spin Kitaev model in materials such as Zr$_{0.5}$Ru$_{0.5}$Cl$_3$.  By developing a superexchange theory specifically for this mixed-spin system, we identify the conditions under which dominant Kitaev-like interactions emerge. Our findings highlight the importance of spin-orbital couplings and quadrupolar order parameters in stabilizing exotic phases, providing a foundation for exploring mixed-spin Kitaev magnets.   
\end{abstract}

\maketitle
\section{Introduction}

The search for quantum spin liquids (QSLs) has been a major focus of condensed matter physics as they represent novel quantum phases of matter characterized by fractionalized excitations, long-ranged quantum entanglement, and emergent gauge fields ~\cite{Anderson1973, Kitaev2006, Balents2010, Savary2016, knolle2019field, Zhou2017, Broholm2020, Takagi2019, Trebst2022}. Among various models, the Kitaev honeycomb model (KHM)~\cite{Kitaev2006} 
has emerged as a paradigmatic Hamiltonian hosting different types of QSLs. Its hallmark is the exact solvability for spin $1/2$, where the model supports fractionalized excitations in the form of Majorana fermions coupled to conserved plaquette fluxes described by a static $Z_2$ gauge field~\cite{Kitaev2006}. Although exact solvability is lost for extensions of the KHM to higher spins ($S > 1/2$), several studies have demonstrated that such systems can still realize $Z_2$ QSLs by identifying an exact representation of the conserved plaquette fluxes using static $Z_2$ gauge fields and employing Majorana fermion representations suitable for larger spins~\cite{Baskaran2008PRB, Rousochatzakis2018NC, Jin2022, Natori2023, Carvalho2023, HanMa2023}. This approach allows systematic analysis via parton mean-field theory, preserving conceptual connections to the exactly solvable spin-$1/2$ case.

Among these higher-spin extensions,
the $S=3/2$ case stands out due to its distinct quantum fluctuations driven by multipolar spin operators, setting it qualitatively apart from the $S=1/2$ model.
These differences manifest in unique spin-liquid instabilities and quantum proximity phases~\cite{Georgiou2024}. Remarkably, for the $S=3/2$ case, density-matrix renormalization group (DMRG) simulations display exceptional quantitative agreement with predictions from the SO(6) Majorana mean-field theory~\cite{Jin2022}, a feature that can be traced back to the underlying model and order-parameter symmetries~\cite{Natori2023}.
 Earlier studies have also shown that while
the  isotropic $S=3/2$ model realizes a quantum spin-orbital liquid (QSOL), small anisotropic deviations of the exchange interactions induce strong first-order transitions into distinct QSL phases characterized by the coexistence of quadrupolar order and either gapped or gapless Majorana excitations~\cite{Natori2023}. Further insights have been obtained by introducing a single-ion anisotropy (SIA) that couples directly to the quadrupolar parameter, highlighting its role as an essential tuning parameter: in the limit of large SIA, the $S=3/2$ KHM effectively reduces to the simpler $S=1/2$ case, thereby providing a controlled pathway between these two regimes~\cite{Natori2023}.

The KHM is also remarkable for its experimental relevance ~\cite{Knolle2017, Takagi2019, Trebst2022, Rousochatzakis2024}, a fertile research field started by Jackeli's and Khaliullin's seminal paper on Mott insulators formed by edge-sharing octahedra of ligands involving  heavy magnetic ions arranged on a honeycomb lattice ~\cite{Jackeli2009, Chaloupka2010}. The Kitaev interaction is particularly relevant to 4$d$ and 5$d$ transition metal compounds, where strong spin-orbit coupling (SOC) leads to effective angular momenta that interact among themselves through highly anisotropic and bond-dependent spin exchanges ~\cite{Khaliullin2005, Jackeli2009, Chaloupka2010, Rousochatzakis2024}. One of the most extensively studied candidate materials is the spin-orbit-coupled Mott insulator $\alpha$-RuCl$_3$, which is believed to host Kitaev interactions and be in close proximity to a QSL state~\cite{Plumb2014, banerjee_neutron_2017, do_majorana_2017, jansa_observation_2018, Takagi2019, Trebst2022}. 

Kitaev-like interactions have been explored in various materials with effective large-$S$ degrees of freedom, driven either by the strong spin-orbit coupling of ligands  or by heavy transition metal magnetic ions 
\cite{Xu2020, Lee2020, Stavropoulos2019, Stavropoulos2021, Yamada2018, Yamada2021, Natori2018, Rousochatzakis2024, Churchill_arXiv2024}.
In particular,
 Yamada \textit{et al.} \cite{Yamada2018, Yamada2021} proposed that exotic QSOL phases could be realized in $\alpha$-ZrCl$_3$  \cite{Swaroop1964, Swaroop1964-2}, a 4$d$ material sharing the same honeycomb lattice structure as $\alpha$-RuCl$_3$. 
 A key distinction in the magnetism of these two materials lies in their electronic configurations. The Zr-based compound features one electron in  the $t_{2g}$ orbital manifold as opposed to one hole in the case of 
 $\alpha$-RuCl$_3$. This difference leads to a $j=3/2$ effective model for $\alpha$-ZrCl$_3$, involving anisotropic and bond-dependent multipolar exchanges \cite{Natori2018,Chen2010}.

In this work, we investigate the mixed-spin KHM, where spin-1/2 and spin-3/2 ions occupy the A and B honeycomb sublattices. Such mixed spin systems are known from the early days of the theory of Mott insulators, i.e., giving rise to the phenomenon of ferrimagnetism \cite{Néel_1952,Wolf_1961}, but the realizing of such QSLs has not been addressed before. Mixing spin-1/2 and spin-3/2 sites within the Kitaev lattice can stabilize unique quantum phases absent in homogeneous systems.  For example, introducing a spin-3/2 defect site into a spin-1/2 KHM can act as a magnetic impurity~\cite{Masahiro2024}, leading to local flux binding effects. These significantly modify the low-energy excitations which offers a new perspective on impurity physics in QSL~\cite{Masahiro2024}. Here, we study a homogeneous lattice of mixed spin 1/2 with spin 3/2, which could be potentially realized experimentally in materials such as Zr$_{0.5}$Ru$_{0.5}$Cl$_3$ and gives rise to a rich phase diagram with entangled spin and orbital degrees of freedom.  


The main results and structure of this paper are as follows:
 Section \ref{sec:MixSp} discusses the exact properties of the mixed-spin KHM, using them to motivate a parton mean-field theory allowing us to sketch its phase diagram. We first discuss the conserved plaquette operators and propose a reformulation of the model in terms of pseudospin and pseudo-orbital operators in Section \ref{sec:conserved}. This enables the use of SO(6) Majorana partons to map conserved $Z_2$ fluxes onto static $Z_2$ gauge fields, facilitating the mean-field analysis in the zero-flux sector that is detailed in Section \ref{sec:MFT}. At the mean-field level, four distinct quantum spin liquid (QSL) phases are identified and summarized in the phase diagram shown in Fig. \ref{fig:phase-diagram}.
Section \ref{sec:DMRG} describes the DMRG approach employed in this study to compute the quadrupolar parameters that differentiate the mean-field QSL phases.  In analogy to Ref. \cite{Jin2022}, we also provide a direct comparison between numerical and analytical results. These two complementary approaches display a remarkable quantitative agreement for most of the phase diagram, with the exception of the region near the isotropic point. In Section \ref{sec:Derivation}, we use the standard strong-coupling approach to derive the superexchange theory for Zr$_{0.5}$Ru$_{0.5}$Cl$_3$, a natural candidate for the mixed-spin Kitaev model. By varying the parameters of an underlying multi-orbital Hubbard model, we identified the necessary conditions for forming dominant Kitaev interactions. The technical details are provided in Appendix \ref{App:superexchange}.
Finally, Section \ref{sec:conclusions} discusses the broader significance of our work and highlights potential directions for future research.

\section{Mixed-spin Kitaev Honeycomb Model}\label{sec:MixSp}

Inspired by the exact solution of the spin-1/2 KHM \cite{Kitaev2006} and the existing theoretical analysis of the spin-3/2 KHM \cite{Jin2022, Natori2023}, we will study the phase diagram of the mixed-spin Kitaev model under a single-ion anisotropy (SIA), which is formally written by 
\begin{equation}
H=\sum_{\lelagm}K_\gm S_{i}^{\gm}J_{j}^{\gm} + H_{\text{SIA}}.\label{eq:model_plus_SIA}
\end{equation}
Here, $S_{i}^{\gm}$ ($J_{i}^{\gm}$) represents a spin-1/2 (spin-3/2) operator, and $\gm$ simultaneously labels the quantization axes and the distinct bond directions. The spin-1/2 ions will be on the sublattice denoted by $A$, whereas spin-3/2 will be on $B$ sublattice. By convention, a unit cell contains two neighboring ions connected by a $z$-bond parallel to the $y$-axis. 

We will also introduce a control parameter in the form of the simplest SIA term 
\begin{equation}
H_{\text{SIA}}=D_z\sum_{j} \left(J_{j}^{z}\right)^{2},\label{eq:SIA}
\end{equation}
which lifts the four-fold degeneracy of  $J = 3/2$ quadruplet by splitting $J_z = \pm 3/2$ and $J_z = \pm 1/2$ levels. As we shall see, this addition allows for a more manageable investigation of the mixed-spin KHM by projecting it to the spin-1/2 model, thus ensuring an exactly solvable limit.

\subsection{Exact results and conserved fluxes}\label{sec:conserved}

The local symmetries of Eq.~\eqref{eq:model_plus_SIA} can be made more transparent in terms of $J=3/2$ pseudospin and pseudo-orbital operators \citep{Jin2022,Natori2023}. We introduce the pseudospins by
\begin{equation}
\sg_{l}^{\gm}=-i\exp\left(i\pi J_{l}^{\gm}\right), \label{eq:pseudoS}
\end{equation}
which corresponds to the local operators forming the spin-$S$ conserved quantities \citep{Baskaran2008PRB}. Likewise, the pseudo-orbital operators read 
\begin{align}
& T_{j}^{z}=\left(J_{j}^{z}\right)^{2}-5/4,\nonumber \\
& T_{j}^{x}=\frac{1}{\sqrt{3}}\left[\left(J_{j}^{x}\right)^{2}-\left(J_{j}^{y}\right)^{2}\right] ,\nonumber \\
&T_{j}^{y}=\frac{2\sqrt{3}}{9}\overline{J_{j}^{x}J_{j}^{y}J_{j}^{z}},\label{eq:pseudoT}
\end{align}
where the bar indicates the sum over all possible permutations. The operators $\left(T^z,T^x \right)$ are quadrupoles that transform as $e_g$ orbital operators under real-space rotations. By contrast, the octupolar operator $T^y$ forms a one-dimensional representation of the $O_h$ symmetry group  \cite{Chen2010}. In conjunction, $\bsym{\sg}$ and $\bf{T}$ satisfy the algebra 
\begin{align}
\left[\sg_{i}^{\al},\sg_{j}^{\bt}\right] & =2i\dt_{ij}\ep^{\al\bt\gm}\sg_{i}^{\gm},\nonumber \\
\left[T_{i}^{\al},T_{j}^{\bt}\right] & =2i\dt_{ij}\ep^{\al\bt\gm}T_{i}^{\gm},\nonumber \\
\left\{ \sg_{i}^{\al},\sg_{j}^{\bt}\right\}  & =\left\{ T_{i}^{\al},T_{j}^{\bt}\right\} =2\dt_{ij}\dt^{\al\bt},\nonumber \\
\left[\sg_{i}^{\al},T_{j}^{\bt}\right] & =0.\label{eq:algebra_sigma_T}
\end{align}
The fifteen operators $\left\{\sg^a, T^b, \sg^a T^b\right\}$ correspond to generators of SU(4), implying that they can be used to rewrite any Hermitian $J=3/2$ operator. A simple example is Eq. (\ref{eq:SIA}), which reads
 \begin{equation} 
 H_{\text{SIA}} = D_z \sum_j T_j^z + \text{const}.\label{eq:SIA_Tz}  
 \end{equation}
We can also write $J_i^{\gm}$ as \cite{Jin2022, Natori2023}
\begin{equation}
J_i^{\gm} = -\frac{\sg_i^{\gm}}{2} - \sg_i^{\gm} T_i^{\al \bt},
\end{equation}
where the compass-like pseudo-orbitals $T^{\al\bt}$ are
\begin{align}
&T_i^{xy} = T_i^z, \quad
T_i^{yz(zx)}= -\frac{T_i^z}{2} \pm \frac{\sqrt{3}T_i^x}{2}.
\end{align}

The  mixed-spin exchange model in Eq.~(\ref{eq:model_plus_SIA}) is then rewritten as 
  \begin{equation}
  H_{\text{K}} = \sumgm K_{\gm}  S_{i}^{\gm} \sg_{j}^{\gm}  \left( \frac{1}{2} + T_j^{\al \bt} \right). \label{eq:mixedKHM_sg_T}
\end{equation}
Crucially, it allows us to show that the Hamiltonian commutes with the local plaquette operators  
\begin{equation}
    W_p = 2^{3}S^{z}_{1}\sg^{x}_{2}S^{y}_{3}\sg^{z}_{4}S^{x}_{5}\sg^{y}_{6},\label{eq:Wp}
\end{equation}
in which the quantization axes correspond to the outward bond label and the multiplying factor $2^3$ was introduced to ensure the $W_p$ eigenvalues are $\pm 1$. The presence of a conserved plaquette flux is in close analogy with the spin-1/2 \citep{Kitaev2006} and spin-3/2 KHM \citep{Baskaran2008PRB, Jin2022, Natori2023}.  The extensive number of conserved quantities indicates that the model realizes a Kitaev QSL and is amenable to an analytical treatment, as we will show in the following section. These conserved quantities also provide guidelines for interpreting the DMRG results, as explored in Section \ref{sec:DMRG}.

The SIA term in Eq.~\eqref{eq:SIA} and $W_p$ also commute, as directly observed from the commutation relations between $T^z$ and $\bsym{\sg}$, allowing us to use it as a control parameter. In the $D_z \rightarrow +\infty$ limit, pseudo-orbital fluctuations are effectively suppressed, allowing us to fix $T^{z} = -1$ and to project the $J^{\gamma}$ angular momenta onto the pseudospins according to the rule:
\begin{equation}
\left(J_{j}^{x},J_{j}^{y},J_{j}^{z}\right)\overset{D_{z}\rightarrow+\infty}{\longrightarrow}\left(-2\sg^{x},-2\sg^{y},\sg^{z}\right).
\end{equation}
Therefore, Eq.~(\ref{eq:model_plus_SIA}) reduces to  
\begin{equation}
  \lim_{D_{z}\rightarrow\infty}H = \sumgm K_{\gm}^\pr S_{i}^{\gm} \sg_{j}^{\gm}.
\end{equation}
with modified coupling constants $\left(K_{x}^{\pr},K_{y}^{\pr},K_{z}^{\pr}\right)=\left(-2K_{x},-2K_{y},K_{z}\right)$. Fixing $T^z=-1$ implies that $\sg^\gm$ will act on the $J^z=\pm 1/2$ manifold only, allowing the pseudospins to be treated as effective  spin-1/2 operators. Thus, the SIA  connects the mixed-spin KHM model to the standard spin-1/2 KHM, therefore introducing an exactly solvable limit.
 
\subsection{Majorana representation and parton mean-field theory}\label{sec:MFT}
An analytical treatment of the mixed-spin KHM
becomes possible by introducing Majorana partons, which map $Z_2$ conserved quantities in Eq. (\ref{eq:Wp}) onto fluxes of a static $Z_2$ gauge field. We begin by representing the spin-1/2 degrees of freedom using the Kitaev Majorana parton framework \citep{Kitaev2006}:
\begin{equation}
    S_{i}^{\gm} =-\frac{i}{2}\eta_{i}^{\gm}c_{i}, \label{eq:S_one-half}
\end{equation}
in which the Majorana particles with flavors $\eta^{\al}$ and $c$ at sites $i,j$ satisfy
\begin{align}
\left\{ \eta_{i}^{\al},\eta_{j}^{\bt}\right\} &   =2\dt_{ij}\dt^{\al\bt},\nonumber \\
\left\{ c_{i},c_{j}\right\} &   =2\dt_{ij},\nonumber \\
\left\{ \eta_{i}^{\al},c_{j}\right\} &   =0\label{eq:Majorana_algebra}
\end{align}
Eq. (\ref{eq:S_one-half}) adequately reproduces the canonical spin-1/2 algebra once we introduce a local gauge operator \cite{Kitaev2006}. Likewise, we introduce the SO(6) Majorana representation for spin-3/2 as follows \citep{Jin2022, Natori2023,Wang2009}
\begin{align}
\bsym{\sg}_{i}=-\frac{i}{2}\bsym{\eta}_{i}\times\bsym{\eta}_{i}, & \,\mathbf{T}_{i} =-\frac{i}{2}\bsym{\ta}_{i}\times\bsym{\ta}_{i},\nonumber\\ \sg_{i}^{\al}T_{i}^{\bt} &=-i\eta_{i}^{\al}\ta_{i}^{\bt},\label{eq:Majorana_parton}
\end{align}
in which the Majorana flavors $\eta$ satisfy the same algebra as in Eq. \ref{eq:Majorana_algebra} plus the added relations with $\theta$ Majorana fermions
\begin{align}
\left\{ \theta_{i}^{\al},\theta_{j}^{\bt}\right\} &   =2\dt_{ij}\dt^{\al\bt},\nonumber \\
\left\{ \eta_{i}^{\al},\theta_{j}^{\bt}\right\} &   = \left\{ c_{i},\theta_{j}^{\bt}\right\}= 0.\label{eq:Majorana_algebra2}
\end{align} 
The first two equations in Eq. \ref{eq:Majorana_parton} correspond to the SO(3) Majorana representation of spin-1/2 systems \citep{Coleman1994,Fu2018} and reproduce the algebra from Eq.~\eqref{eq:algebra_sigma_T}. The second line allows us to represent all spin-orbital operators as bilinears and is consistent with the constraint on the physical Hilbert space \cite{Wang2009}
\begin{equation}
{\mathcal D}_{i}=i\eta_{i}^{\al}\eta_{i}^{\bt}\eta_{i}^{\gm}\ta_{i}^{\al}\ta_{i}^{\bt}\ta_{i}^{\gm}=1.\label{eq:projector}
\end{equation}
The ${\mathcal D}_i$  operator not only enforces the constraint but also provides an alternative representation of the pseudospin operators as quartic operators, as discussed in recent studies \citep{Jin2022,Natori2023,Schaden2023}:
\begin{equation}
    \sg_{j}^{\gm}=-\frac{i}{2}\eta_{j}^{\gm}\ta_{j}^{0},\label{eq:alt_sigma}
\end{equation}
where $\ta_{j}^{0}\equiv-i\ta_{j}^{x}\ta_{j}^{y}\ta_{j}^{z}$. This allows us to reproduce the algebra of all $S=3/2$ operators in terms of Majorana bilinears.

The Hamiltonian in Eq.~(\ref{eq:model_plus_SIA}), expressed in the Majorana parton representation, takes the form:
\begin{align}
    H &=\frac{1}{4}\sumgm K_{\gm}\hat{u}_{\lelagm}ic_{i}\left(\ta_{j}^{0}+2\ta_{j}^{\al\bt}\right)\nonumber\\
    & - D_z \sum_j i \ta_j^{x} \ta_j^{y},\label{eq:MajRepH}
\end{align}
in which $\hat{u}_{\lelagm}=-i\eta_{i}^{\gm}\eta_{j}^{\gm}$ represents a static $Z_2$ gauge operator. The replacement of $\left\{\hat{u}_{\lelagm} \right\}$ by their $\pm1$ eigenvalues determines the flux sectors labeled by $W_p$. The zero-flux sector ($W_p = +1,\forall p$)  is the ground state of the  Kitaev spin-1/2 model, as established by Lieb's theorem \citep{Lieb1994}. Furthermore, both numerical and analytical studies provide strong evidence that this sector also hosts the ground state of the KHM for arbitrary spin-$S$ \citep{Baskaran2008PRB, Rousochatzakis2018NC, Jin2022, Natori2023}. Thus, we will focus on the zero-flux sector of the mixed-spin KHM,  a choice further supported by the DMRG simulations presented later in this work.
 
The Hamiltonian in Eq. (\ref{eq:MajRepH}), after gauge fixing, consists of Majorana bilinear terms with additional quartic interactions of the form $i c_i \ta_{j}^{0}$, which require a mean-field treatment. The most general mean-field decoupling within a fixed gauge sector can be expressed as:
\begin{align}
    ic_{i}\ta_{j}^{0}&\approx\Delta_{\lelagm}^{x}i\ta_{j}^{y}\ta_{j}^{z}+\Delta_{\lelagm}^{y}i\ta_{j}^{z}\ta_{j}^{x}+\Delta_{\lelagm}^{z}i\ta_{j}^{x}\ta_{j}^{y}\nonumber\\&+Q_{j}^{x}ic_{i}\ta_{j}^{x}+Q_{j}^{y}ic_{i}\ta_{j}^{y}+Q_{j}^{z}ic_{i}\theta_{j}^{z},
\end{align}
with the parameters defined by
\begin{align}
 \Delta_{\lelagm}^{t} &=-\left\langle \Psi_0 \left| ic_i\ta_{j}^{t} \right| \Psi_0 \right\rangle \nonumber\\
 Q_{j}^{t} &=\left\langle T_{j}^{t} \right\rangle =-\left\langle \Psi_0 \left| i\ta_{j}^{r}\ta_{j}^{s} \right| \Psi_0 \right\rangle, \label{eq:selfconpar}
\end{align} 
in which $\left\langle  \Psi_0 \left| \hat{O} \right|  \Psi_0 \right\rangle$ is the expectation value of $\hat{O}$ for the mean-field ground state $\left|\Psi_0\right\rangle$. The $\Delta_{\lelagm}^{t}$ parameters correspond to the pairings $c_i$ and $\ta_{j}^{t}$, which are Majorana matter fermions from spin-1/2 and spin-3/2, respectively. On the other hand, nonvanishing $Q_{j}^{t}$ accounts for the possible coexistence of a QSL with a quadrupolar or octupolar parameter \cite{Jin2022, Natori2023}. This is expected to be the case when introducing Eq. \ref{eq:SIA_Tz} in the limit $D_z\rightarrow \infty$, where $Q^{z}\rightarrow -1$ uniformly throughout the lattice. 

We will enforce the zero-flux sector by fixing $u_{\lelagm}=+1$, leading to the following mean-field decoupled Hamiltonian then reads:
\begin{align}
H_{\text{MFT}} & =\frac{1}{2}\sum_{\mb{r},\gm}K_{\gm}ic_{\mb{r}}\ta_{\mb{r}_{\gm}}^{\al\bt}\nonumber \\
 & +\frac{1}{4}\sum_{\mb{r},\gm,t}K_{\gm}Q^{t}ic_{\mb{r}}\ta_{\mb{r}_{\gm}}^{t}\nonumber \\
 & +\frac{1}{4}\sum_{\mb{r},\gm,t}K_{\gm}\Delta_{\gm}^{t}i\ta_{\mb{r}_{\gm}}^{r}\ta_{\mb{r}_{\gm}}^{s}\nonumber \\
 & -D_{z}\sum_{\mb{r}_{\gm}}i\ta_{\mb{r}_{\gm}}^{x}\ta_{\mb{r}_{\gm}}^{y},\label{eq:MFT}
\end{align} 
in which $\mb{r}$ label spin-1/2 ions and $\mb{r}_{\gm}$ their spin-3/2 nearest neighbors. We applied a sum convention in the third line by fixing $\epsilon^{rst}=1$. We also presupposed translationally invariant mean-field parameters, thus making their position labels superfluous. Our mean-field theory will then depend upon 12 parameters, three of them being onsite ($Q^t$) and nine corresponding to pairings involving $\ta^t$ operators along $\gm$-bonds ($\Delta_{\gm}^t$). For computation purposes, writing the Majorana fermions in terms of canonical fermions $f$ and $g$ will be conveniently defined as follows

\begin{align}
c_{\mb{r},A}=f_{\mb{r}}+f_{\mb{r}}^{\dg}, & \,\ta_{\mb{r},B}^z=g_{\mb{r}}^{\dg}+g_{\mb{r}},\nonumber \\
 \ta_{\mb{r},B}^y=\frac{f_{\mb{r}}-f_{\mb{r}}^{\dg}}{i},& \,\ta_{\mb{r},B}^x=\frac{g_{\mb{r}}-g_{\mb{r}}^{\dg}}{i}.
\end{align} 
Eq. (\ref{eq:MFT}) in the Fourier space can be written in terms of a $4\times4$ matrix $H_{\mb{k}}$ as follows
\begin{align}
H_{\text{MFT}}&=\sum_{\mb{k}}\Psi_{\mb{k}}^{\dg}H_{\mb{k}}\Psi_{\mb{k}}\nonumber\\
&=\sum_{\mb{k}}\Phi_{\mb{k}}^{\dg}(U_{\mb{k}}^{\dg}H_{\mb{k}}U_{\mb{k}})\Phi_{\mb{k}}\nonumber\\ &\equiv\sum_{\mb{k}}\Phi_{\mb{k}}^{\dg}\Omega_{\mb{k}}\Phi_{\mb{k}},
\end{align}
where  $\Psi_{\mb{k}}^{\dagger} = \left(f_{\mb{k}}^\dg, g_{\mb{k}}^\dg, f_{-\mb{k}}, g_{-\mb{k}} \right)$, $\Phi_{\mb{k}}^{\dg} = \Psi_{\mb{k}}^{\dg}U_{\mb{k}}$ are the eigenstates, and $\Omega_{\mb{k}} =U_{\mb{k}}^{\dg} H_{\mb{k}} U_{\mb{k}}$ is the diagonal matrix of eigenvalues. The mean-field ground state $\left|\Psi_{0}\right\rangle$ satisfies  $\Phi_{\mb{k}}\left|\Psi_{0}\right\rangle = 0$ for all $\Phi_{\mb{k}}$ corresponding to negative eigenstates. 

The self-consistency algorithm begins with an initial guess $\mb{x}_0= \left(\Delta_{\gm}^{x},..,Q^{z}\right)$ for the order parameters. This initial guess defines the starting Hamiltonian, the unitary matrices $U_{\mb{k}}$, as well as the ground state $\left|\Psi_{0}(\mb{x}_0)\right\rangle$. The initial guess also allows us to evaluate a vector $\mb{x}_\text{par}(\mb{x}_0)$ using Eq. (\ref{eq:selfconpar}). The self-consistent condition can then be reformulated as a root-finding problem for the function $f(\mb{x})=\mb{x}_\text{par}(\mb{x})-\mb{x}$. Such problems are efficiently solved using standard algorithms such as Broyden's method \citep{Ralko2020}. For concreteness, we considered self-consistent solutions $\left|f(\mb{x})\right|< \text{tol}$, in which the tolerance was of the order $10^{-12}$. The algorithm was run with $\sim200$ initial guesses $\mb{x}_0$ to ensure that all self-consistent solutions could be found.     

\begin{figure}
\begin{centering}
\includegraphics[width=0.50\columnwidth]{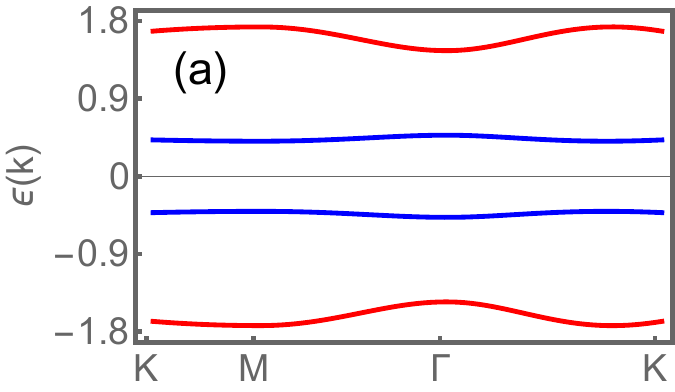}\includegraphics[width=0.46\columnwidth]{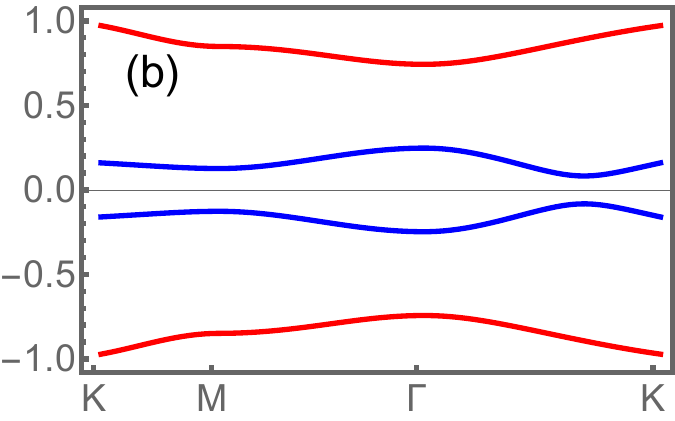}\par\end{centering}
\begin{centering}
\includegraphics[width=0.48\columnwidth]{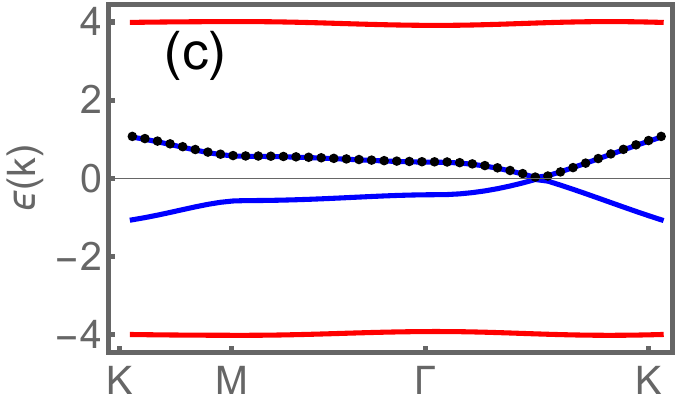}\includegraphics[width=0.46\columnwidth]{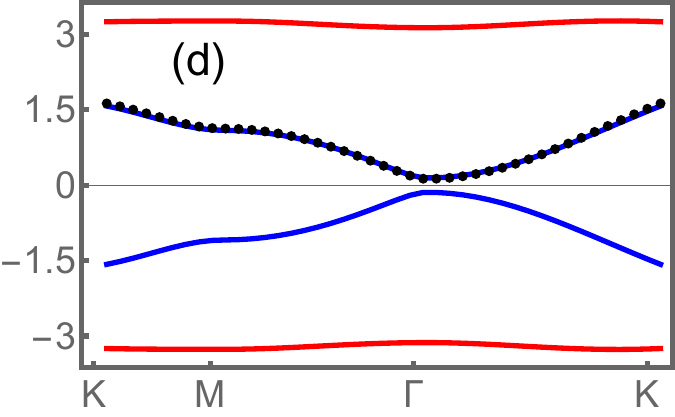} \par\end{centering}
\caption{\label{fig:mean-field-spectrum} Dispersion of the Majorana bands for representative mean-field ground states obtained for fixed $K_{x}=K_{y}=1$. (a) $K_{z}=2$, $D_{z}=0$, representing the gapped $B$ phase with
$Q^{z}>0$, $Q^{x}=0$, (b) $K_{z}=0.5$, $D_{z}=0$ representing
the twofold degenerate $C$ phase with $Q^{z}<0$, $Q^{x}\protect\neq0$, (c) $K_{z}=2.3$, $D_{z}=4$, representing a spin liquid with $Q^{z}\approx-1$
adiabatically connected to the gapless Kitaev spin liquid $A_{0}$, (d) $K_{z}=4.5$, $D_{z}=4$ represents a spin liquid with $Q^{z}\approx-1$ adiabatically connected to the gapped Kitaev spin liquid $A_{z}$. The black points on the figure corresponds to the exact Majorana dispersion of the spin-1/2 KHM with modified coupling constants  $\left(K_{x}^{\prime},K_{y}^{\prime},K_{z}^{\prime}\right)=\left(-2K_{x},-2K_{y},K_{z}\right)$. }
\end{figure}

The mean-field ground state is determined by the self-consistent solution $\mb{x}$ that minimizes the ground-state energy:
\begin{equation}
E_{\text{GS}}=\sum_{\mb{k},\Omega_{\mb{k},a}<0}\Omega_{\mb{k},a}.    
\end{equation} 
Fig. \ref{fig:mean-field-spectrum} illustrates some representative dispersions for fixed coupling constants  $K_{x}=K_{y}=1$ but varying $K_{z}$ and $D_{z}$ values. For all the parameters examined, we found  $Q^y = 0$, which implies that time-reversal invariance is preserved. Figs. \ref{fig:mean-field-spectrum}(a) and \ref{fig:mean-field-spectrum}(b) display gapped spin liquids in the small single-ion anisotropy limit (in this case, we set $D_{z}=0$) that are physically distinguished by the order parameters $Q^{z}$ and $Q^{x}$. Fig. \ref{fig:mean-field-spectrum}(a) displays a $K_{z}>1$ case, for which the gapped liquid $B$ is unique and characterized by $Q^{z}>0$ and $Q^{x}=0$. This is analogous to the toric-code phase, being described in terms of dimers on the $z-$bonds containing $S^{z}=\pm3/2$ states. Fig. \ref{fig:mean-field-spectrum}(b) displays the dispersion of a twofold degenerate gapped liquid $C$ that is stabilized when $K_{z}<1$. This phase is characterized by
$Q^{z}<0$ and $Q_{x}\neq0$, where  $Q_{x}$ can take two values of equal magnitude but opposite sign.

The $B$ and $C$ KSLs are degenerate at the isotropic point $K_z = 1$ and $D_z = 0$, implying a threefold degenerate mean-field ground state. More explicitly, this solution is characterized by  
\begin{eqnarray}
(Q^z, Q^x)_\text{MFT}&=&0.845(1,0)  \nonumber\\
(Q^z, Q^x)_\text{MFT}&=&0.845\left(-\frac{1}{2},\pm \frac{\sqrt{3}}{2}\right).   
\label{eq:QzQX_iso_MFT}\end{eqnarray}
This result should be compared with the spin-3/2 KHM, for which this point is critical and displays vanishing quadrupolar parameters \cite{Jin2022,Natori2023}. By contrast, the isotropic mixed-spin KHM is characterized by a level crossing between two possible mean-field states. A comparison between this result and the one obtained through DMRG will be given in the next section. 

\begin{figure}
\begin{centering}
\includegraphics[width=0.9\columnwidth]{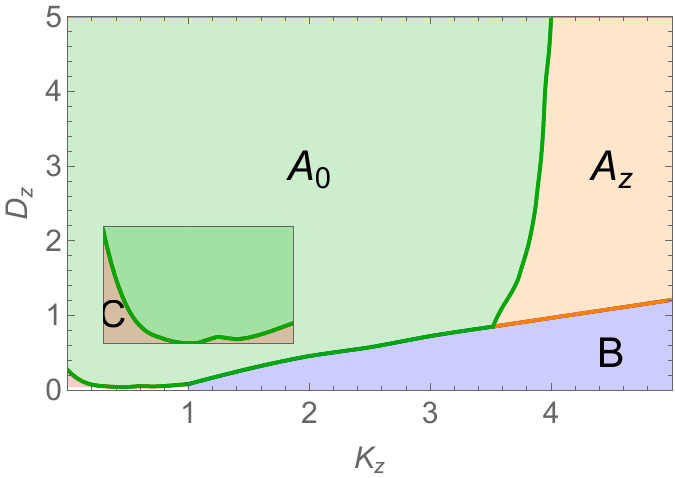}
\par\end{centering}
\caption{\label{fig:phase-diagram} Mean-field phase diagram showing the quantum spin liquids hosted by the mixed-spin Kitaev model. The phases $A_{0}$, $A_{z}$, and $B$ occupy the largest area of the diagram. The phase $C$ occupies a small area of the diagram with small $D_{z}$ and $0<K_{z}<1$, as shown on the inset. }
\end{figure}

The introduction of SIA connects the mixed-spin and the $S=1/2$ KHM,  providing a regime in which the mean-field theory recovers known exact results.  Figs. \ref{fig:mean-field-spectrum} (c) and (d)  illustrate two cases in the large SIA limit, where the parameter $Q^{z}$ is very close to -1.  In this limit, both the gapless ($A_{0}$) and gapped ($A_{z}$) phases  are adiabatically connected to the $S=1/2$ Kitaev QSL. This connection is explicitly demonstrated by the agreement between the mean-field low-energy bands and the exact Majorana fermion dispersion of the  $S=1/2$ KHM with appropriately modified coupling constants \cite{Kitaev2006}. The full phase diagram of the model (\ref{eq:model_plus_SIA})  is displayed in Fig. \ref{fig:phase-diagram}. It reveals the dominance of the
 $B$, $A_0$, and $A_z$ phases  across most of the parameter space, with the $C$ phase occupying only a narrow region when $0<K_z<1$.

\section{Numerical simulations}\label{sec:DMRG}

To examine the validity and robustness of our parton mean-field theory, we perform state-of-the-art density matrix renormalization group (DMRG) simulations ~\cite{White1992, White1993} to investigate the ground state of Hamiltonian ~\eqref{eq:model_plus_SIA}. These calculations are performed on a two-dimensional honeycomb lattice comprising $L_y \times L_x$ unit cells, arranged in a cylindrical geometry. Periodic boundary conditions (PBC) are applied along the shorter dimension (circumference $L_x$), while the longer dimension (length $L_y$) remains open. This cylindrical setup explicitly breaks the $C_3$ rotational symmetry of the lattice. To ensure high numerical accuracy, we use a bond dimension of up to $\chi=4000$, achieving a typical truncation error of approximately $\epsilon \simeq 10^{-6}$.

The ground states obtained by DMRG simulations exhibit a zero-flux configuration in both the $A_0$ and $A_z$ phases in accordance with our Ansatz. In the $B$ phase, however, our DMRG simulations do not converge to a unique ground-state flux configuration. Instead, they yield a disordered-flux state, where the flux on each plaquette deviates from the expected values of $1$ or $-1$. This behavior is reminiscent of findings in a previous study on the $S=3/2$ Kitaev honeycomb model~\cite{Jin2022}, where this phenomenon was attributed to  an extremely small energy gap associated with $Z_2$ flux flipping in the $B$ phase.

\begin{figure}
\begin{centering}
\includegraphics[width=0.8\columnwidth]{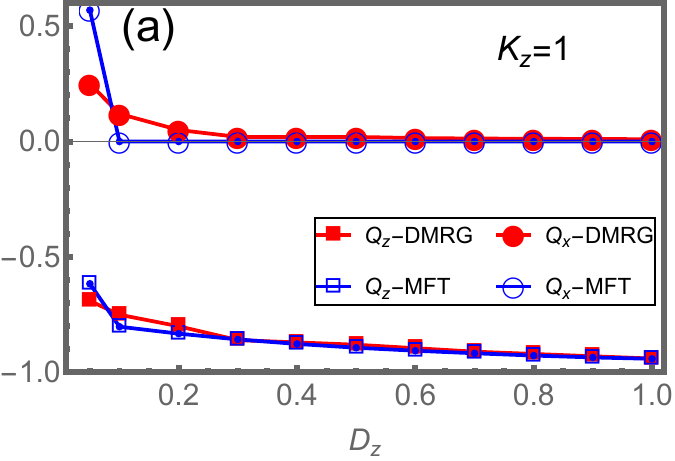}
\par\end{centering}
\begin{centering}
\includegraphics[width=0.8\columnwidth]{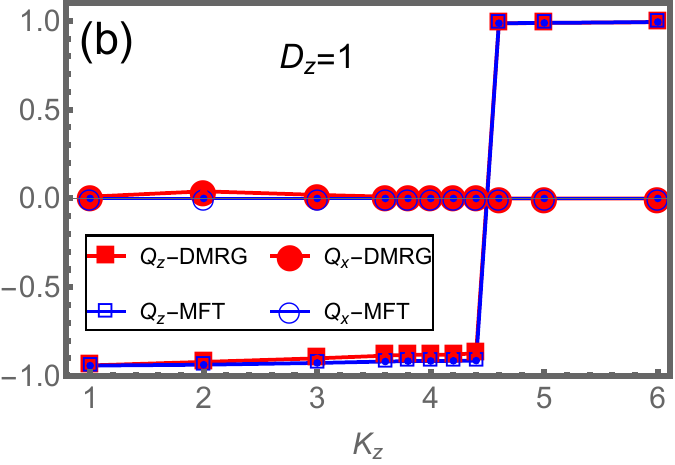}
\par\end{centering}
\caption{\label{fig:comparison} Comparison between computed order parameters using parton mean-field theory (in blue and open markers) and DMRG (in red). The $Q^z$ order parameter is indicated with squares, while $Q^x$ is displayed with circles. Fig. (a) investigates the parameters with fixed exchange $K_z = 1$ but varying $D_z$. In (b), the single-ion anisotropy is fixed at $D_z = 1$, but the $K_z$ is varied over values such }
\end{figure}

We also calculate the averaged expectation values of $J=3/2$ multipole operators $Q^z=\langle T^z\rangle$ and $Q^x=\langle T^x\rangle$. These two values are spatially uniform in the bulk of cylinders and are qualitatively consistent with the parton mean-field theory, as demonstrated by the data in Table~\ref{tab:DMRGR} and Fig. \ref{fig:comparison}.

\begin{table}
\caption{ Averaged expectation values of $Q^z=\langle{}T^z\rangle$ and $Q^x=\langle{}T^x\rangle$, as well as ground-state flux configurations. The data is obtained on a cylinder with $L_y=4$ and $L_x=12$.}\label{tab:Z2DSL}~\label{tab:DMRGR}
\begin{tabular}{ p{2cm} p{1.5cm} p{1.5cm} p{1.5cm} }
    \hline
    \hline
    \rule{0pt}{2.5ex}    
    ($K_z$, $D_z$)   &  $Q^z$ & $Q^x$ &   zero-flux \\
    \hline
    (1.1, 0.0)   &   0.84    &  -1e-5 & No \\
    (1.05,0.0)   &   0.82    & -5e-5  & No \\
    (1.0, 0.05)  &  -0.69    & 0.25   & Yes \\
    (1.0, 0.1)   &  -0.75    & 0.12   & Yes \\
    (1.0, 0.2)   &  -0.8     & 0.05   & Yes \\
    (1.0, 0.3)   & -0.86     & 0.02   & Yes \\
    (1.0, 1.0)   &   -0.94   &  0.01  & Yes \\
    (3.0, 1.0)   &   -0.90   &  0.02  & Yes \\             
    (3.6, 1.0)   &   -0.89   &  0.01  & Yes \\
    (3.8, 1.0)   &   -0.88   &  0.01  & Yes \\
    (4.4, 1.0)   &   -0.87   &  0.01  & Yes \\
    (4.6, 1.0)   &   0.985   & -2e-5  & No \\ 
    (5.0, 1.0)   &   0.988   &  1e-7  & No \\
    \hline
    \hline
\end{tabular}
\end{table}

\begin{figure}
    \centering
    \includegraphics[width=0.6\linewidth]{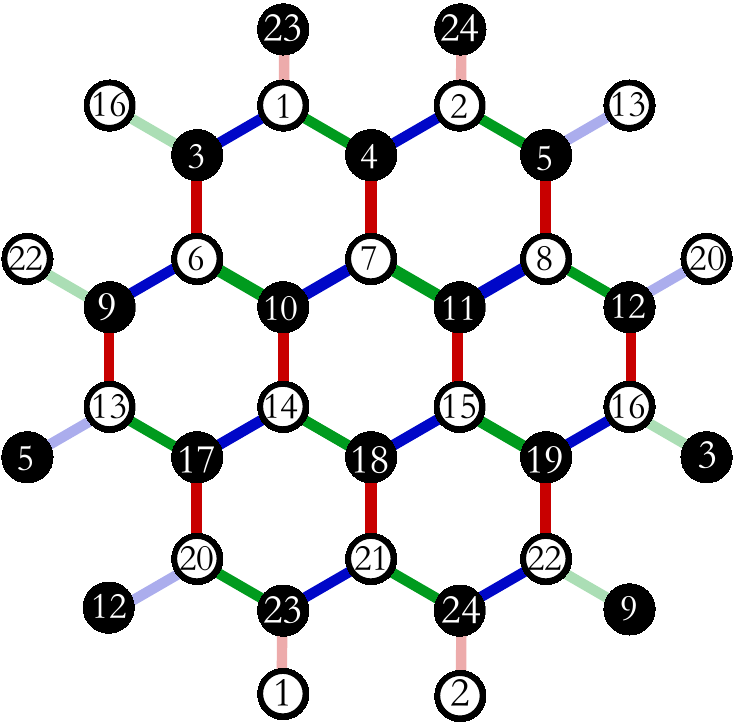}
    \caption{A 24 lattice-site cluster with $C_3$ rotational symmetry. The white and black dots the $S=1/2$ and $J=3/2$ spins, respectively. The blue, green, and red bonds represent the $x-$, $y-$, and $z-$type bonds. The transparent bonds denote the boundary bonds.}
    \label{fig:24site}
\end{figure}

For the isotropic point of $K_z=1$ and $D_z=0$,  we use exact diagonalization to calculate the ground state on a torus. To preserve the $C_3$ rotational symmetry, we consider a 24 lattice-site cluster shown in Fig.~\ref{fig:24site}.  Note that this cluster geometry breaks the translation symmetries in both directions. We implement the $Z_2$ flux conservation (e.g., a local symmetry) with the QuSpin package~\cite{quspin}. Focusing on the zero-flux sector, we find that the ground states exhibit a 3-fold degeneracy for the $C_3$ rotational symmetry in which the expectation values of $T^z$ and $T^x$ operators manifest exactly the same relative values as those predicted by the parton mean-field theory, namely,

\begin{eqnarray}
(Q^z, Q^x)_\text{ED}&=&(-0.1214,0) \nonumber \\
(Q^z, Q^x)_\text{ED}&=&(+0.0607,0.1051) \\
(Q^z, Q^x)_\text{ED}&=&(+0.0607,-0.1051)\nonumber
\label{eq:QzQX_MFT}
\end{eqnarray}
Moreover, we find that the first excited states in the zero-flux sector also display the same 3-fold degeneracy. These results are consistent with those obtained with DMRG at the same point but calculated on a more well-defined ground state.

Comparing the DMRG  results with the mean-field results at the isotropic point Eq.(\ref{eq:QzQX_iso_MFT}) shows a qualitative agreement with respect to the relative values of $Q^x$ and $Q^z$, but a strong quantitative disagreement. We explore the nature of this disagreement in Fig. \ref{fig:comparison}(a) by first fixing the value $K_z = 1$ and varying the SIA. Starting at large values of $D_z$, we observe a strong quantitative agreement between the two techniques for both order parameters up to $D_z \sim 0.1$. Below this range, there is a sizable divergence between the computed parameters, specially for $Q^x$. Fig. \ref{fig:comparison}(b) indicates a complementary analysis in which $D_z = 1$ is fixed in order to ensure only the $A_0$ and $A_z$ phases. The quantitative agreement between the evaluated parameters is recovered, even concerning the location of the phase transition. 

Our analysis indicates that mean-field theory and DMRG will converge to the same kind of spin liquids except in the neighborhood of the isotropic point, namely, parton mean-field theory converges to the $B$ or $C$ phases that are continuously connected to the same spin liquids at $K_z > 1$ and $K_z<1$. On the other hand, DMRG and exact diagonalization predict a sign inversion of the $Q^z$ order parameter leading to a qualitatively different spin liquid in this region. Furthermore, this parameter will display a reduced absolute value, but not a vanishing one as observed for the spin-3/2 KHM \citep{Jin2022,Natori2023}, a feature that cannot be explained by $C_3$ symmetry constraints \citep{Natori2023}. The nature of the isotropic mixed-spin Kitaev spin liquid also differs from the large-$S$ Kitaev spin liquids \citep{Rousochatzakis2018NC}, since the quadrupolar parameters preserve translational symmetry. Thus, the mixed-spin model stabilizes a qualitatively different QSL, whose nature is not yet tractable within our parton Ansatz.

\section{Derivation of the mixed spin superexchange Hamiltonian}\label{sec:Derivation}

The identification of solid-state platforms exhibiting Kitaev interactions has established a vibrant research direction at the intersection of materials science and strongly correlated electron systems~\cite{Takagi2019, Trebst2022, Knolle2017, Winter2016PRB}. Within this context, an important question is whether ferrimagnetic Kitaev spin liquids can be realized in systems with mixed spin magnitudes. A promising scenario involves a honeycomb lattice composed of $j = 1/2$ and $j = 3/2$ ions, where anisotropic, bond-dependent exchanges are expected to be significant. 

\begin{figure}[t]
	\centering
        \includegraphics[width=\linewidth]{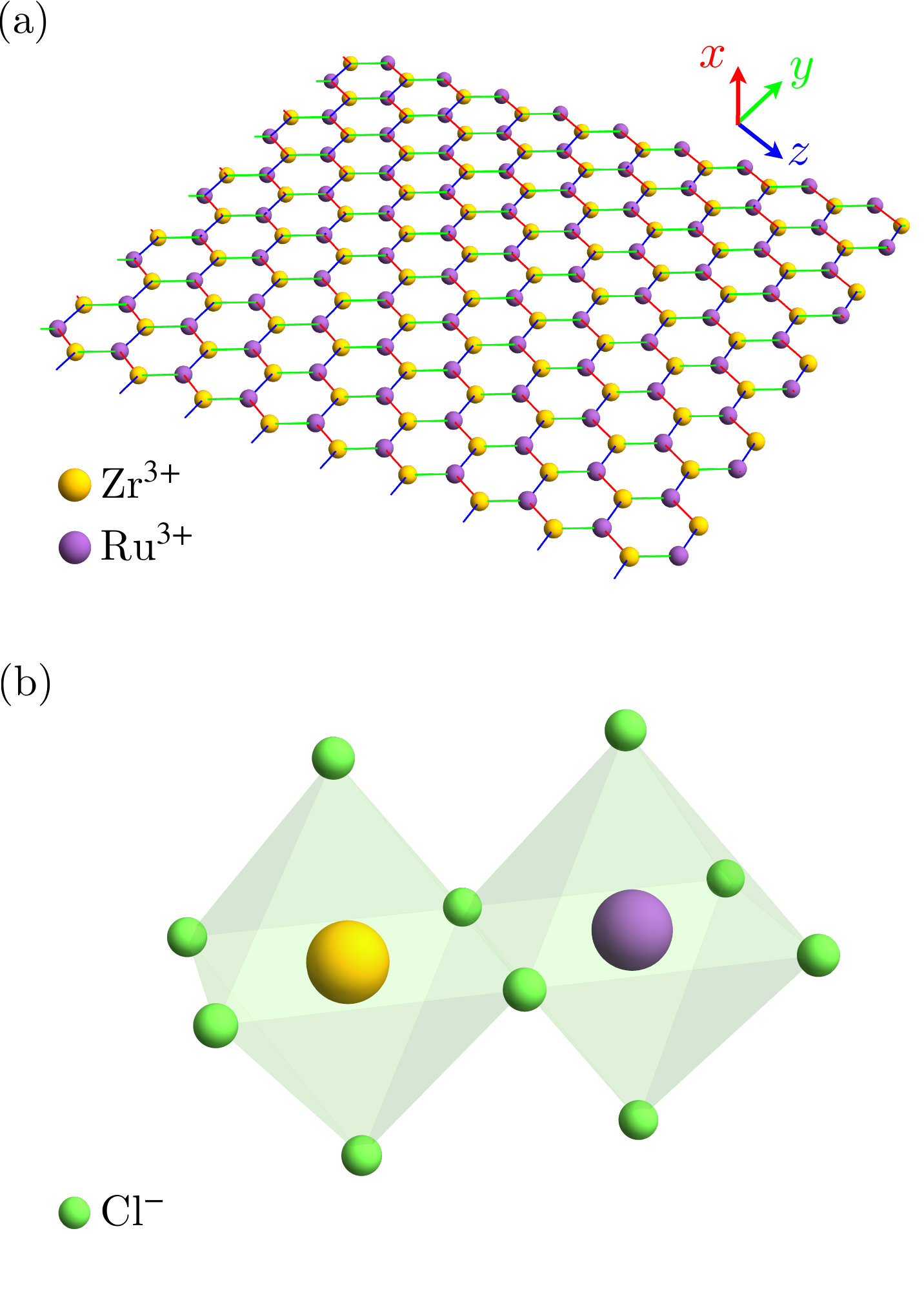}
        \caption{(a) The mixed spin-1/2 and spin-3/2 Kitaev model, where Zr${}^{3+}$ and Ru${}^{3+}$ ions form a honeycomb lattice (ligand Cl${}^{-}$ ions are not shown) connected by $x$, $y$, and $z$ bonds (red, green, blue) perpendicular to $x$, $y$, and $z$ axis, respectively. (b) The local environment of a bond connected by Zr${}^{3+}$ and Ru${}^{3+}$ ions, where Zr${}^{3+}$ and Ru${}^{3+}$ sit in the center of their own octahedral crystal field cage formed by Cl$^{-}$ ions. }\label{fig:lattice}
\end{figure}

In this section, we derive the minimal microscopic model for Zr$_{0.5}$Ru$_{0.5}$Cl$_3$, a candidate compound featuring Ru$^{3+}$ ions in a $d^5$ configuration and Zr$^{3+}$ ions in a $d^1$ configuration.   These ions are arranged according to Fig.~\ref{fig:lattice} (a) and each is surrounded by six Cl$^-$ ions.
 We focus on the undistorted lattice geometry, where the crystal field environment splits the fivefold degenerate $d$ orbitals into lower-energy $t_{2g}$ and higher-energy $e_g$ levels. Assuming a large crystal-field splitting, the relevant low-energy physics arises from electrons confined to the $t_{2g}$ manifold. Accordingly, all local interactions, including spin-orbit coupling, Coulomb repulsion, and Hund’s coupling, are treated within this subspace.

The microscopic Hamiltonian describing this hybrid magnetic ion system can be derived from a three-band Hubbard model, which accounts for the electronic interactions within the $t_{2g}$ orbitals of both Ru$^{3+}$ and Zr$^{3+}$ ions. This approach incorporates the effects of on-site Coulomb interactions, Hund's coupling, spin-orbit coupling, and hopping processes mediated by the intermediate Cl$^-$ ligands. It is given by
\begin{align}
    \mathcal{H}_{t_{2g}}=\mathcal{H}_{\mathrm{ion}}+\mathcal{H}_t,
\end{align}
where $\mathcal{H}_t$ gives the hopping between the $t_{2g}$ orbitals on Zr and Ru ions, and  the single-ion Hamiltonian  given by
 \begin{widetext}
     \begin{align}
            \mathcal{H}_{\mathrm{ion}}=\sum_{i}&\left[U_1\sum_{\alpha}n_{i\alpha\uparrow}n_{i\alpha\downarrow}+\frac{1}{2}(U_2-J_H)\sum_{\alpha\neq\alpha',\sigma}n_{i\alpha\sigma}n_{i\alpha'\sigma}+U_2\sum_{\alpha\neq\alpha'}n_{i\alpha\uparrow}n_{i\alpha'\downarrow}\right.\nonumber\\+&\left.J_H\sum_{\alpha\neq\alpha'}d_{i\alpha\uparrow}^\dagger d_{i\alpha\downarrow}^\dagger d_{i\alpha'\downarrow}d_{i\alpha'\uparrow}-J_H\sum_{\alpha\neq\alpha'}d_{i\alpha\uparrow}^\dagger d_{i\alpha\downarrow}d_{i\alpha'\downarrow}^\dagger d_{i\alpha'\uparrow}+\mathcal{H}_{i,\mathrm{SOC}}\right],
    \end{align}       
\end{widetext}
where
$d_{i \alpha \sigma}^\dagger$ denotes the creation operator of the $d$-electron on the magnetic ion $i$ on the $t_{2g}$ orbital $\alpha = xy,\, yz,\, zx$  with spin $\sigma = \uparrow, \downarrow$,
$n_{i\alpha\sigma}$  is the corresponding number operator. The term 
 $\mathcal{H}_{i,\mathrm{SOC}}$ represents the spin-orbit coupling on site $i$, characterized by the strength $\lambda$.
 The constants $U_1$ and $U_2$ denote the Coulomb repulsion
among d-electrons on the same and on the different $t_{2g}$ orbitals,
respectively, $J_H$ denotes the Hund's coupling constant, and
 and the intraorbital repulsion satisfies $U_1=U_2+2J_H$.
Since Zr and Ru are close in the periodic table, we use the same set of $U_2$, $J_H$, $\lambda$ for both Zr and Ru. 

Noting that the energy of the single-ion Hamiltonian is dominated by the number of electrons occupying the $t_{2g}$ orbital, it would energetically favorable to transfer one electron from the Ru$^{3+}$ ion to the Zr$^{3+}$ ion to minimize the energy of the single-ion Hamiltonian. To avoid this situation, which is magnetically inert, we introduce a positive onsite potential energy $V$ on the Zr$^{3+}$ ions to stabilize the ground state of the Zr--Ru system in the desired $d^{1}$--$d^{5}$ configuration. We assume $V$ to be large enough that the $d^1$--$d^5$ configuration will be the ground state of the system, but not so large as to favor the $d^0$--$d^6$ configuration. The acceptable range for $V$ is given by: $\frac{1}{4}(12U_2-6J_H-3\lambda+\sqrt{16J_H^2+8 J_H \lambda +9\lambda^2}+2\sqrt{25J_H^2+10J_H \lambda+9\lambda^2})<V<5U_2+\frac{3\lambda}{2}$. For reasonable parameters, such as  $U_2=2.0$ eV, $J_H=0.4$ eV, and  $\lambda=0.15$ eV, we find that $7~\mathrm{eV}\lesssim V \lesssim 10~\mathrm{eV}$. 

The stabilized $d^{1}$–$d^{5}$ electronic configuration results in a ground-state manifold described by the effective spin states $|J^z_{J=3/2}, J^z_{J=1/2}\rangle$. Assuming that the mixed spin-1/2 and spin-3/2 system remains insulating, we model the virtual electron hoppings using $\mathcal{H}_t$. By treating the hopping of electrons between the $t_{2g}$ orbitals as a perturbation, we derive the superexchange Hamiltonian in the basis of  $|J^z_{J=3/2}, J^z_{J=1/2}\rangle$, which takes the form:
\begin{widetext}
\begin{align}\label{eq:SEij}
        \mathcal{H}_{\mathrm{eff},ij}=\sum_{n,m}^8\sum_{\mathrm{excited}}\frac{\langle n_i|\mathcal{H}_t| \mathrm{excited}\rangle\langle \mathrm{excited}|\mathcal{H}_t|m_j\rangle}{E_{\mathrm{Zr}}^{(0)}+E_{\mathrm{Ru}}^{(0)}-(E_{\mathrm{excited}}\pm V)}|n_i\rangle\langle m_j|,
\end{align}
\end{widetext}
where $n_i$, $m_j$ denotes $n$-th, $m$-th state from the 
$|J^z_{J=3/2} J^z_{J=1/2}\rangle$ basis on site $i$ and $j$ respectively, $E_{\mathrm{Zr}}^{(0)}$, $E_{\mathrm{Ru}}^{(0)}$ are the ground state energies of the single-ion Hamiltonian, and $E_{\mathrm{excited}}\pm V$ 
corresponds to the energy of the excited state with the sign $\pm$ depending on the excited electronic configuration, $d^2$--$d^4$ or $d^0$--$d^6$.

The resulting superexchange Hamiltonian on the $z$ bond is expressed as a product of orthogonal spin-3/2 and spin-1/2 operators, with the corresponding coupling strengths detailed in Table~\ref{tab:coupling1}. This formulation captures the anisotropic nature of the interactions arising from the underlying spin-orbital coupling and crystal field effects. In addition to these interactions, the perturbative calculations also introduce SIA terms for the spin-3/2 degrees of freedom, as shown in Table~\ref{tab:coupling2} in the form of $Q_i$ terms. However, since we are keeping the $C_3$ rotational symmetry of the honeycomb lattice, the terms $Q_1$ and $Q_2$ sum to zero when contributions from all three bonds are considered. While the SIA terms $Q_3$ and $Q_4$ contribute non-zero terms, they remain subdominant across the entire range of hopping parameters considered, which is why they are not shown here. We will mostly focus on the $D_i$, representing dipole-dipole interactions between spin-1/2 and spin-3/2 moments, and the  $O_i$ terms correspond to the couplings associated with the higher-order interactions.

 
 

\begin{figure*}     
     \centering
            \includegraphics[width=1\textwidth]{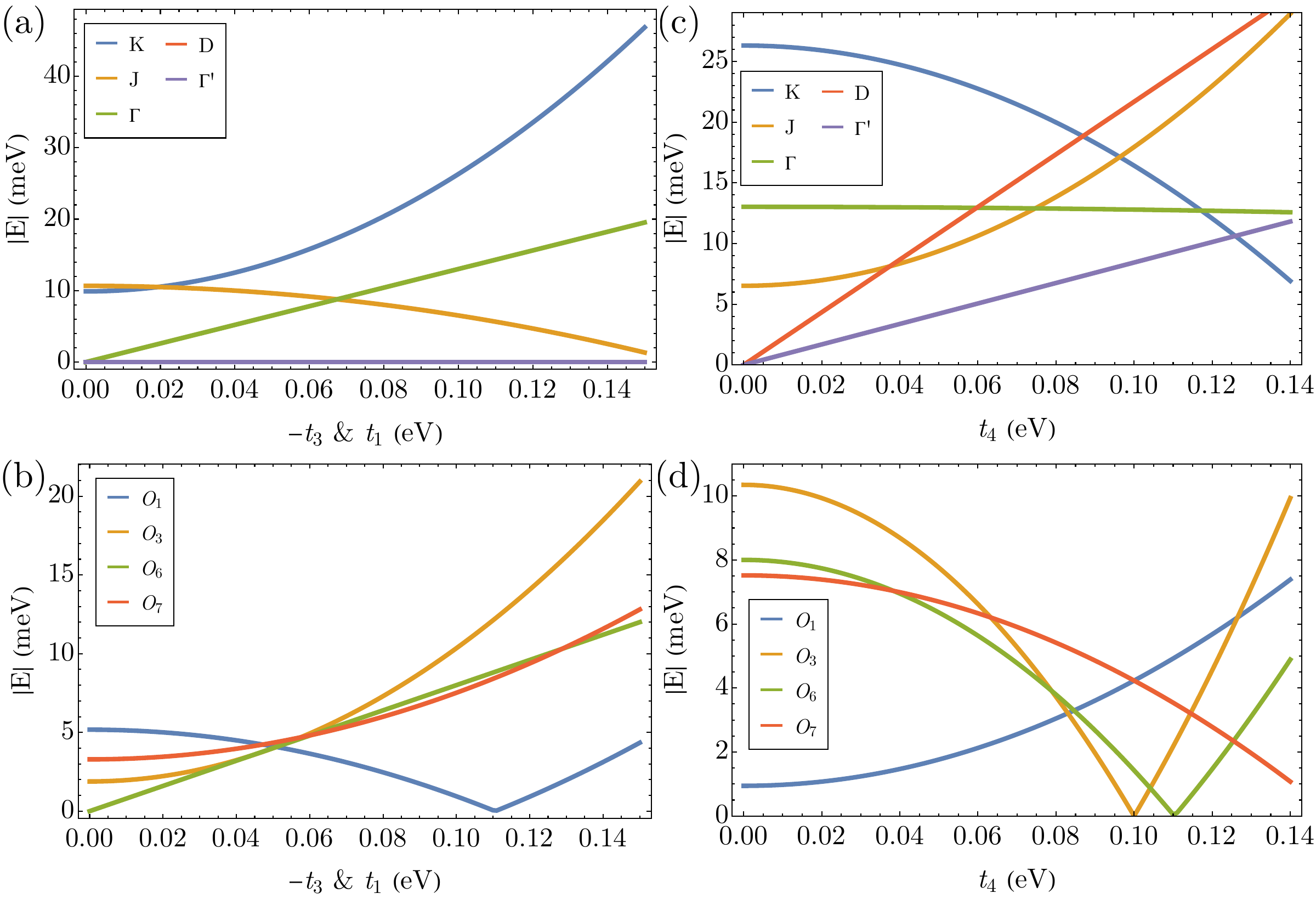}
            \caption{  Superexchange interactions for the mixed spin-1/2–spin-3/2 system. The parameters are fixed as $U_2 = 2$ eV, $J_H = 0.4$ eV, and $\lambda = 0.15$ eV. In panels (a) and (c), we show the dipole-dipole couplings $J$, $K$, $\Gamma$, $\Gamma'$, and $D$. In panels (b) and (d), the largest higher-order interactions, $O_1$, $O_3$, $O_6$, and $O_7$, are presented. For panels (a) and (b), the hopping parameters are set as $t_2 = 0.114$ eV and $t_4 = 0$ eV, while $t_1 = -t_3$ is varied from 0 to 0.15 eV. In panels (c) and (d), we fix $t_2 = 0.114$ eV and $t_1 = -t_3 = 0.1$ eV, and vary $t_4$ from 0 to 0.15 eV.
           }\label{fig:couplings}
        \end{figure*}     
\begin{table}[h]
\caption{Superexchange coupling constants between spin-3/2 and spin-1/2 degrees of freedom on the $z$ bond.}\label{tab:coupling1}
    \centering
    \begin{tabular}{|C|C|C|C|}
                 \hline
                         & S_x  & S_y& S_z\\\hline 
                 J_x     & D_1 & D_2 & D_4\\\hline
                 J_y     & D_2 & D_1 & D_4\\\hline
                 J_z     & D_4' & D_4' & D_3\\\hline
                 (J^x)^3 & O_1 & O_2 & O_4\\\hline
                 (J^y)^3 & O_2 & O_1 & O_4\\\hline
                 (J^z)^3 & O_4' & O_4' & O_3\\\hline
                 J^xJ^yJ^z+J^zJ^yJ^x & O_5 & O_5 & O_6\\\hline
                 J^xJ^zJ^z-J^yJ^yJ^x & O_7 & O_8 & O_{9}\\\hline
                 J^yJ^zJ^z-J^xJ^xJ^y & O_8 & O_7 & O_{9}\\\hline
                 J^yJ^yJ^z-J^zJ^xJ^x & O_{10}&-O_{10}& O_{11}\\\hline
    \end{tabular}
\end{table}

\begin{table}[h]
    \caption{Single ion anisotropy for spin-3/2 degrees of freedom induced by the hopping on the $z$ bond.}\label{tab:coupling2}
    \centering
    \begin{tabular}{|C|C|}
                \hline
                -2J^xJ^x+J^yJ^y+J^zJ^z &Q_1\\\hline
                J^zJ^z-J^yJ^y &Q_2\\\hline
                J^xJ^y+J^yJ^x &Q_3\\\hline
                J^xJ^z+J^zJ^x &Q_4\\\hline
                J^yJ^z+J^zJ^y &Q_4\\\hline
    \end{tabular}.
\end{table}
Focusing on the dipole-dipole couplings between the spin-1/2 and spin-3/2 moments, we rewrite these matrices in a familiar format commonly used for Kitaev materials. In this notation, we assign $J = D_1$, $K = D_3$, $\Gamma = D_2$, and $\Gamma' + D = D_4$ for the dipole-dipole couplings, and $\Gamma' - D= D_4'$ for other interactions. Here $D$ denotes the Dzyaloshinskii-Moriya interaction. With these definitions, the effective Hamiltonian on the $z$-bond can be expressed as:
\begin{eqnarray}
\label{eq:JKGGpmodelij}
\renewcommand{\arraystretch}{1.5}
\begin{array}{l}
\mathcal{H}_{ij,z}\!=\!J {\bf S}_i\!\cdot\!{\bf J}_j\!+\!K S^z_i J^z_j
\!+\!\Gamma (S^{x}_iJ^{y}_j\!+\!S^{y}_iJ^{x}_j)\\
~~~~~~+\Gamma'\big(S^{x }_i J^z_j + S^z_iJ^{x}_j+S^{y}_i J^z_j + S^z_iJ^{y}_j\big)\\
~~~~~~+D\big(S^{x }_i J^z_j - S^z_iJ^{x}_j+S^{y}_i J^z_j - S^z_iJ^{y}_j\big) + ...\,,
\end{array}
\end{eqnarray}
where $...$ includes terms for higher-order interactions and single-ion anisotropy contributions. 

We first numerically examine these coupling constants by using parameters typically associated with $\alpha$-RuCl${}_{3}$: $U_2 = 2$ eV, $J_H = 0.4$ eV, $\lambda = 0.15$ eV, and the hopping parameters $t_1 = 0.066$ eV, $t_2 = 0.114$ eV, $t_3 = -0.229$ eV, and $t_4 = -0.010$ eV. We also assume an onsite potential of $V = 8$ eV for the Zr ions. 
It provides an estimate for the strength of the superexchange coupling constants, which are presented in Table \ref{tab:coupling3} of Appendix \ref{App:superexchange}.   While these parameters give a dominant Kitaev interaction, there are still sizable contributions from the non-Kitaev exchanges. 
 
 To explore whether it is possible to further maximize the Kitaev interaction while suppressing non-Kitaev terms, we systematically vary these parameters in order to identify an optimal set that enhances the Kitaev interaction and ensures it dominates over competing terms. Since the dominatant direct hopping terms are $t_1$ and $t_3$ given the local geometry, we fix all parameters to their $\alpha$-RuCl${}_{3}$ values but set  $t_4 = 0$ eV and impose $t_1 = -t_3$. We then vary $t_1$ from 0 to 0.15 eV to explore the parameter space for maximizing the Kitaev interaction. The resulting couplings are shown in  Fig.~\ref{fig:couplings} (a) and (b). We observe that for $t_1 = -t_3$ in the range of 0.04 to 0.1 eV, the Kitaev interaction is the largest coupling, ranging from 12 to 20 meV. However, other interactions remain significant. For instance, the $\Gamma$ interaction is approximately half the strength of the Kitaev interaction, and several higher-order terms, such as $O_3$, $O_6$, and $O_7$, also contribute significantly to the overall coupling landscape.

Next, we fix $t_2 = 0.114$ eV and $t_1 = -t_3 = 0.1$ eV, and vary $t_4$ from 0 to 0.15 eV. The resulting couplings are shown in Fig.~\ref{fig:couplings} (c) and (d). Our results show that increasing the  $t_4$ hopping reduces the Kitaev interaction while simultaneously increasing the Heisenberg coupling $J$ and, in particular, the Dzyaloshinskii-Moriya interaction $D$.

We conclude this section by noting that a more realistic evaluation of the microscopic parameters will require the use of advanced ab-initio and quantum chemistry methods.
 While the on-site interaction parameters $U_1$, $U_2$, and $J_H$  are expected to be of comparable magnitude in both
  Ru and Zr ions, they will generally differ due to the distinct electronic configurations, resulting in different exchange couplings. Similarly, assessing whether the intersite interaction parameter $V$ lies within a physically reasonable range  requires more quantitative methods.
Furthermore, in any mixed spin-1/2–spin-3/2 system, the hopping amplitudes are expected to deviate from those in $\alpha$-RuCl$_3$ due to the presence of different spin species and possible lattice distortions. Such distortions can lift the $C_3$ symmetry of the honeycomb lattice, thereby permitting single-ion anisotropy (SIA) terms like those introduced in Eq.~\ref{eq:SIA}.
Despite these simplifications, our perturbative analysis serves as a proof of principle demonstrating that a mixed-spin Kitaev honeycomb model can arise as an effective description of ferrimagnetic systems, introducing exchange frustration into a setting long studied from a different perspective. A detailed exploration of the full phase diagram and refinement of the minimal model are left for future work.



\section{Conclusions}\label{sec:conclusions}

In this paper, we explored the mixed-spin Kitaev honeycomb model, where alternating spin-1/2 and spin-3/2 ions occupy the two sublattice positions of the honeycomb lattice. We constructed a comprehensive phase diagram using parton mean-field theory and DMRG simulations. The phase diagram reveals four distinct QSL phases, each characterized by a unique quadrupolar order parameter, specific flux configurations, and Majorana fermion excitations. The quantitative agreement between parton mean-field analysis and numerical approaches highlights the robustness of our framework, except in a small parameter region around the isotropic point. We also discussed a potential experimental realization in materials such as Zr$_{0.5}$Ru$_{0.5}$Cl$_3$, in which we derived a microscopic superexchange Hamiltonian and identified  conditions under which dominant Kitaev-like interactions arise.   

Our results foster further investigation of frustrated magnetism in ferrimagnetic systems. While our study focused on pure Kitaev interactions with single-ion anisotropy,  taking into account other interactions in Eq.(\ref{eq:JKGGpmodelij}), such as the Heisenberg exchange,  bond-anisotropic $\Gamma$-interaction,  and higher-order multipolar couplings, could stabilize other exotic phases, including chiral QSLs and magnetically ordered states. These extensions provide further opportunities for future research to explore the interplay between dipolar and multipolar interactions, potentially uncovering an even broader spectrum of quantum phases.

  \vspace*{0.3cm} 
\noindent{\it  Acknowledgments:} 
We thank  Onur Erten, Wen-Han Kao, Masahiro Takahashi, Rodrigo Pereira, and Eric Andrade for  useful discussions.
The work by N.B.P. was supported by the  National Science Foundation under Award No.\ DMR-1929311.   
N.B.P. 
acknowledges the hospitality and partial support of the Technical University of Munich – Institute for Advanced Study and the support of the Alexander von Humboldt Foundation. N.B.P. and J.K.  also thank the hospitality
of  Aspen Center for Physics,
which is supported by National Science Foundation
grant PHY-2210452.
JK acknowledges support from the Deutsche Forschungsgemeinschaft (DFG, German Research Foundation) under Germany’s Excellence Strategy– EXC–2111–390814868 and DFG Grants No. KN1254/1-2, KN1254/2-1 and TRR 360 - 492547816, as well as the Munich Quantum Valley, which is supported by the Bavarian state government with funds from the Hightech Agenda Bayern Plus. J.K. further acknowledges support from the Imperial-TUM flagship partnership.
Y.Y. was supported by the US Department of Energy Basic Energy Sciences under Contract No. DE-SC0020330.


\appendix
\setcounter{figure}{0}
\setcounter{equation}{0}
\renewcommand{\theequation}{A\arabic{equation}}
\renewcommand{\thefigure}{A\arabic{figure}}


\section{Microscopic derivation of the superexchange Hamiltonian} \label{App:superexchange} 
 In this Appendix  we provide  details on the microscopic derivation of the superexchange Hamiltonian.  Section \ref{sec:one-particle} presents a comprehensive single-ion description of Zr and Ru ions, identifying the local microscopic parameters and the single-ion eigenstates that define the local degrees of freedom. 
 Section \ref{sec:hop} discusses the physical origins of the relevant hopping integrals and establishes the notation used throughout.
 Section \ref{sec:perturbation}   outlines 
 the perturbation expansion, including the classification of virtual states and their  energies.
Section \ref{sec:model_derivation} details the projection of the perturbation matrix onto a set of orthogonal spin matrices, expressing the superexchange Hamiltonian in terms of the corresponding spin operators. This section also clarifies the specific set of spin-3/2 operators used in the projection.

\subsection{One-particle eigenstates}\label{sec:one-particle}
The spin-orbit coupling (SOC) interaction couples the spin $S = 1/2$ of either the single hole in Ru$^{3+}$ or the single electron in Zr$^{3+}$ to their effective orbital angular momentum $L = 1$, resulting in total angular momenta of $J = 1/2$ and $J = 3/2$, respectively. Consequently, for the single electron in Zr$^{3+}$, the lowest-energy state is four-fold degenerate, with an energy of $E_{\mathrm{Zr}}^{(0)} = -\frac{\lambda}{2}$. The corresponding eigenstates are given by:
\begin{align}
        |E_{\mathrm{Zr},1}^{(0)}\rangle &= \frac{1}{\sqrt{2}}\left(-i|d_{yz,\uparrow}\rangle+|d_{xz,\uparrow}\rangle\right),\\
        |E_{\mathrm{Zr},2}^{(0)}\rangle &= \frac{1}{\sqrt{6}}\left( -i|d_{yz,\downarrow}\rangle+|d_{xz,\downarrow}\rangle+2i|d_{xy,\uparrow}\rangle\right),\\
        |E_{\mathrm{Zr},3}^{(0)}\rangle &= \frac{1}{\sqrt{6}}\left(i|d_{yz,\uparrow}\rangle+|d_{xz,\uparrow}\rangle+2i|d_{xy,\downarrow}\rangle\right),\\
        |E_{\mathrm{Zr},4}^{(0)}\rangle &= \frac{1}{\sqrt{2}}\left(i|d_{yz,\downarrow}\rangle+|d_{xz,\downarrow}\rangle\right).
\end{align}
Similarly, the lowest-energy states  of Ru$^{3+}$  have an energy $E_{\mathrm{Ru}}^{(0)} = 10U_2 - \lambda$, and they are given by the following eigenstates:
\begin{align}
        |E_{\mathrm{Ru},1}^{(0)}\rangle=&\frac{1}{\sqrt{3}}\left(|d_{yz,\uparrow}\rangle+i|d_{xz,\downarrow}\rangle+|d_{xy,\downarrow}\rangle\right),\\
            |E_{\mathrm{Ru},2}^{(0)}\rangle=&\frac{1}{\sqrt{3}}\left(-|d_{yz,\downarrow}\rangle-i|d_{xz,\uparrow}\rangle+|d_{xy,\uparrow}\rangle\right).
\end{align} 
The four degenerate ground states for the single electron on Zr$^{3+}$ become the magnetic degrees of freedom for $J_{\mathrm{eff}}=3/2$, and the two degenerate ground states for five electrons on Ru$^{3+}$ become the magnetic degrees of freedom for $J_{\mathrm{eff}}=1/2$. Now we can derive the superexchange Hamiltonian for $J_{\mathrm{eff}}=3/2$ and $J_{\mathrm{eff}}=1/2$ moments as an $8\times 8$ perturbation matrix with the hopping.\\
\subsection{Hopping matrix} \label{sec:hop}
 
 The effective hopping Hamiltonian between sites on the honeycomb lattice occupied by spin-1/2 and spin-3/2 ions reads
\begin{eqnarray}
\mathcal{H}_t=\sum_{ij}\sum_{\alpha\beta\sigma} 
t_{ij}^{\alpha\beta}d^\dagger_{i\alpha\sigma} d_{j\beta\sigma}
,
\end{eqnarray}
where  ${d}_{i\alpha\sigma}$ are the annihilation operators for the $\alpha$-th orbital with spin $\sigma$ ($\uparrow$ or  $\downarrow$) at site $i$, and  $t_{ij}^{\alpha\beta}$ represents the hopping parameters,  which, in  
the most general case,  can be expressed in matrix form for each bond. For the $z$-bond, the hopping matrix is given by \cite{Rau2014}:                      \begin{align*}
            \begin{tabular}{|c|c|c|c|}
            \hline    & $d_{yz,\sigma}$ & $d_{xz,\sigma}$ & $d_{xy,\sigma}$ \\\hline
            $d_{yz,\sigma}$  & $t_1$    & $t_2$    &      $t_4$ \\\hline
            $d_{xz,\sigma}$  & $t_2$    & $t_1$    &      $t_4$ \\\hline
            $d_{xy,\sigma}$  & $t_4$    & $t_4$    &      $t_3$ \\\hline
            \end{tabular}\, .
        \end{align*}
For the ideal octahedra without any trigonal distortion, there is an additional,  {\it local} $C_2$ symmetry around the axis perpendicular to  the bond  ($[001]$ axis for the  $z$-bond) and passing through its center which prevents any mixing between the $xy$ and the $zx$ and $yz$  orbitals, forcing $t_4\!=\!0$. In the presence of the trigonal distortion, we can  have a nonzero $t_4$.  Also note that the indirect hopping through the ligand ion is accounted for by the renormalization of $t_2$. Finally, the corresponding matrices  for the bonds $x$ and $y$ can be found by applying the $C_{3}$ rotation around the $[111]$ axis.
\subsection{Perturbation theory}\label{sec:perturbation}
Using the perturbation expansion for the effective superexchange Hamiltonian Eq.~(\ref{eq:SEij}), we explicitly account for both $i\rightarrow j$ and $j\rightarrow i$ hoppings, as $i$ and $j$ sites are occupied by inequivalent  Ru$^{3+}$ and Zr$^{3+}$ ions. The excited intermediate states resulting from these single-electron hoppings correspond to the  $d^0$--$d^6$ and $d^2$--$d^4$ configurations. The $d^0$--$d^6$ configuration is reached when an electron hops from the $d^1$ to the $d^5$ state, while the $d^2$--$d^4$ configuration occurs when hopping takes place from $d^5$ to $d^1$.
In the case of the $d^0$--$d^6$ configuration, there is only one excited state, with energy $E_{\mathrm{excited},06} = 15U_2$, where the Zr orbitals are empty and the Ru orbitals are fully occupied. The situation is more complex for the $d^2$--$d^4$ configuration.  The $d^2$ configuration on the Zr$^{2+}$ ion gives rise to 15 intermediate states, and similarly, the $d^4$ configuration on the Ru$^{4+}$ ion results in 15 intermediate states.
In this case, 
each Zr$^{2+}$ or  Ru$^{4+}$ ion gives $5$ distinct eigenvalues:
\begin{align}
            &\mathcal{E}_{\mathrm{Zr},1}=\frac{1}{2}(2U_2-2J_H+\lambda),\nonumber\\ 
            &\mathcal{E}_{\mathrm{Zr},(2,3)}=\frac{1}{2}\left(2U_2+3J_H+\lambda\mp\sqrt{25J_H^2-10J_H\lambda+9\lambda^2}\right),\nonumber\\
            &\mathcal{E}_{\mathrm{Zr},(4,5)}=\frac{1}{4}\left(4U_2-\lambda\pm\sqrt{16J_H^2+8J_H\lambda+9\lambda^2}\right),\nonumber\\
            &\mathcal{E}_{\mathrm{Ru},1}=\frac{1}{2}(12U_2-2J_H-\lambda),\nonumber\\
                      &\mathcal{E}_{\mathrm{Ru},(2,3)}=\frac{1}{4}\left(24U_2+\lambda\mp\sqrt{16J_H^2+8J_H\lambda+9\lambda^2}\right),&&\nonumber\\
                       &\mathcal{E}_{\mathrm{Ru},(4,5)}=\frac{1}{2}\left(12U_2+3J_H-\lambda\pm\sqrt{25J_H^2-10J_H\lambda+9\lambda^2}\right).
        \end{align}
Combining the two ions results in a total of 225 intermediate states, corresponding to 25 distinct intermediate energies, $E_{\mathrm{excited},24}$, which originate from the various excited states of the Zr$^{2+}$ and Ru$^{4+}$ ions.

       Finally, we explicitly compute the superexchange Hamiltonian in Eq.~(\ref{eq:SEij})   in the form of an $8\times 8$ matrix by summing over all the intermediate excited states. This process is carried out systematically using Mathematica. 

\subsection{Spin-1/2 - spin-3/2 Hamiltonian}\label{sec:model_derivation}
After constructing the perturbation matrix, we project it onto a set of orthogonal spin matrices to express the superexchange Hamiltonian in terms of the corresponding spin operators. There are multiple representations of the superexchange Hamiltonian, as various orthogonal spin matrices can be employed to describe the spin-3/2 degrees of freedom. We use a specific set of spin-3/2 operators, consisting of 15 distinct operators: $J_x$, $J_y$, $J_z$, $-2J_x^2+J_y^2+J_z^2$, $J_z^2-J_y^2$, $J_xJ_y+J_yJ_x$, $J_xJ_z+J_zJ_x$, $J_yJ_z+J_zJ_y$, $(J^x)^3-\frac{41}{20}J_x$, $(J^y)^3-\frac{41}{20}J_y$, $(J^z)^3-\frac{41}{20}J_z$, $J^xJ^yJ^z+J^zJ^yJ^x$, $J^xJ^zJ^z-J^yJ^yJ^x$, $J^yJ^zJ^z-J^xJ^xJ^y$, $J^yJ^yJ^z-J^zJ^xJ^x$, and the identity matrix \footnote{The octupolar term $(J^x)^3$ in the Hamiltonian is derived from the orthogonal basis component $(J^x)^3-\frac{41}{20}J_x$, and the remaining dipolar part $-\frac{41}{20}J_x$ is absorbed in the dipolar-dipolar interaction, i,e., $J_x S_x$, $J_x S_y$, and $J_x S_z$.}. This set includes both the fundamental angular momentum components $J_x$, $J_y$, $J_z$ and higher-order terms, capturing the full complexity of the spin-3/2 system. For the spin-1/2 degrees of freedom, we use the conventional Pauli matrices $S_x$, $S_y$, and $S_z$, scaled by the spin length of 1/2. The operators $S_x$, $S_y$, $S_z$ and $J_x$, $J_y$, $J_z$ satisfy the commutation relations $\left\{ S_{l}^{\alpha},S_{m}^{\beta}\right\} =i\epsilon^{\alpha\beta\gamma}\delta_{lm}S_{l}^{\gamma}$
and $\left\{ J_{l}^{\alpha},J_{m}^{\beta}\right\} =i\epsilon^{\alpha\beta\gamma}\delta_{lm}J_{l}^{\gamma}$, respectively. The projection of the perturbation matrix gives us the superexchange Hamiltonian on the $z$-bond shown in Table~\ref{tab:coupling1} and Table~\ref{tab:coupling2} from the main text. The numerical values of the superexchange Hamiltonian assuming parameters for $\alpha$-RuCl${}_{3}$ are shown in Table~\ref{tab:coupling3}.

\begin{table}[h]
    \centering
    \begin{tabular}{|C|C|C|C|}
                 \hline
                         & S_x  & S_y& S_z\\\hline 
                 J_x     & -4.2208 & 25.0595 & -2.24465\\\hline
                 J_y     & 25.0595 & -4.2208 & -2.24465\\\hline
                 J_z     & 4.16899 & 4.16899 & 45.979\\\hline
                 (J^x)^3 & -1.9056 & -6.76715 & 1.02079\\\hline
                 (J^y)^3 & -6.76715 & -1.9056 & 1.02079\\\hline
                 (J^z)^3 & -2.10781 & -2.10781 & -21.9104\\\hline
                 J^xJ^yJ^z+J^zJ^yJ^x & 1.33998 & 1.33998 & 13.8579\\\hline
                 J^xJ^zJ^z-J^yJ^yJ^x & 12.9319 & 7.41444 & -1.98801\\\hline
                 J^yJ^zJ^z-J^xJ^xJ^y & 7.41444 & 12.9319 & -1.98801\\\hline
                 J^yJ^yJ^z-J^zJ^xJ^x & 0.110846 & -0.110846 & 0\\\hline
    \end{tabular}
    \caption{Numerical values of superexchange coupling constants on the $z$ bond assuming parameters for $\alpha$-RuCl${}_{3}$.}\label{tab:coupling3}
\end{table}


\bibliography{references}

\begin{thebibliography}{51}%
\makeatletter
\providecommand \@ifxundefined [1]{%
 \@ifx{#1\undefined}
}%
\providecommand \@ifnum [1]{%
 \ifnum #1\expandafter \@firstoftwo
 \else \expandafter \@secondoftwo
 \fi
}%
\providecommand \@ifx [1]{%
 \ifx #1\expandafter \@firstoftwo
 \else \expandafter \@secondoftwo
 \fi
}%
\providecommand \natexlab [1]{#1}%
\providecommand \enquote  [1]{``#1''}%
\providecommand \bibnamefont  [1]{#1}%
\providecommand \bibfnamefont [1]{#1}%
\providecommand \citenamefont [1]{#1}%
\providecommand \href@noop [0]{\@secondoftwo}%
\providecommand \href [0]{\begingroup \@sanitize@url \@href}%
\providecommand \@href[1]{\@@startlink{#1}\@@href}%
\providecommand \@@href[1]{\endgroup#1\@@endlink}%
\providecommand \@sanitize@url [0]{\catcode `\\12\catcode `\$12\catcode `\&12\catcode `\#12\catcode `\^12\catcode `\_12\catcode `\%12\relax}%
\providecommand \@@startlink[1]{}%
\providecommand \@@endlink[0]{}%
\providecommand \url  [0]{\begingroup\@sanitize@url \@url }%
\providecommand \@url [1]{\endgroup\@href {#1}{\urlprefix }}%
\providecommand \urlprefix  [0]{URL }%
\providecommand \Eprint [0]{\href }%
\providecommand \doibase [0]{https://doi.org/}%
\providecommand \selectlanguage [0]{\@gobble}%
\providecommand \bibinfo  [0]{\@secondoftwo}%
\providecommand \bibfield  [0]{\@secondoftwo}%
\providecommand \translation [1]{[#1]}%
\providecommand \BibitemOpen [0]{}%
\providecommand \bibitemStop [0]{}%
\providecommand \bibitemNoStop [0]{.\EOS\space}%
\providecommand \EOS [0]{\spacefactor3000\relax}%
\providecommand \BibitemShut  [1]{\csname bibitem#1\endcsname}%
\let\auto@bib@innerbib\@empty
\bibitem [{\citenamefont {Anderson}(1973)}]{Anderson1973}%
  \BibitemOpen
  \bibfield  {author} {\bibinfo {author} {\bibfnamefont {P.~W.}\ \bibnamefont {Anderson}},\ }\href@noop {} {\bibfield  {journal} {\bibinfo  {journal} {Materials Research Bulletin}\ }\textbf {\bibinfo {volume} {8}},\ \bibinfo {pages} {153} (\bibinfo {year} {1973})}\BibitemShut {NoStop}%
\bibitem [{\citenamefont {Kitaev}(2006)}]{Kitaev2006}%
  \BibitemOpen
  \bibfield  {author} {\bibinfo {author} {\bibfnamefont {A.}~\bibnamefont {Kitaev}},\ }\href {https://doi.org/10.1016/j.aop.2005.10.005} {\bibfield  {journal} {\bibinfo  {journal} {Annals of Physics}\ }\textbf {\bibinfo {volume} {321}},\ \bibinfo {pages} {2} (\bibinfo {year} {2006})}\BibitemShut {NoStop}%
\bibitem [{\citenamefont {Balents}(2010)}]{Balents2010}%
  \BibitemOpen
  \bibfield  {author} {\bibinfo {author} {\bibfnamefont {L.}~\bibnamefont {Balents}},\ }\href {https://doi.org/10.1038/nature08917} {\bibfield  {journal} {\bibinfo  {journal} {Nature}\ }\textbf {\bibinfo {volume} {464}},\ \bibinfo {pages} {199} (\bibinfo {year} {2010})}\BibitemShut {NoStop}%
\bibitem [{\citenamefont {Savary}\ and\ \citenamefont {Balents}(2017)}]{Savary2016}%
  \BibitemOpen
  \bibfield  {author} {\bibinfo {author} {\bibfnamefont {L.}~\bibnamefont {Savary}}\ and\ \bibinfo {author} {\bibfnamefont {L.}~\bibnamefont {Balents}},\ }\href {http://stacks.iop.org/0034-4885/80/i=1/a=016502} {\bibfield  {journal} {\bibinfo  {journal} {Rep. Prog. Phys.}\ }\textbf {\bibinfo {volume} {80}},\ \bibinfo {pages} {016502} (\bibinfo {year} {2017})}\BibitemShut {NoStop}%
\bibitem [{\citenamefont {Knolle}\ and\ \citenamefont {Moessner}(2019)}]{knolle2019field}%
  \BibitemOpen
  \bibfield  {author} {\bibinfo {author} {\bibfnamefont {J.}~\bibnamefont {Knolle}}\ and\ \bibinfo {author} {\bibfnamefont {R.}~\bibnamefont {Moessner}},\ }\href@noop {} {\bibfield  {journal} {\bibinfo  {journal} {Annual Review of Condensed Matter Physics}\ }\textbf {\bibinfo {volume} {10}},\ \bibinfo {pages} {451} (\bibinfo {year} {2019})}\BibitemShut {NoStop}%
\bibitem [{\citenamefont {Zhou}\ \emph {et~al.}(2017)\citenamefont {Zhou}, \citenamefont {Kanoda},\ and\ \citenamefont {Ng}}]{Zhou2017}%
  \BibitemOpen
  \bibfield  {author} {\bibinfo {author} {\bibfnamefont {Y.}~\bibnamefont {Zhou}}, \bibinfo {author} {\bibfnamefont {K.}~\bibnamefont {Kanoda}},\ and\ \bibinfo {author} {\bibfnamefont {T.-K.}\ \bibnamefont {Ng}},\ }\href {https://doi.org/10.1103/RevModPhys.89.025003} {\bibfield  {journal} {\bibinfo  {journal} {Rev. Mod. Phys.}\ }\textbf {\bibinfo {volume} {89}},\ \bibinfo {pages} {025003} (\bibinfo {year} {2017})}\BibitemShut {NoStop}%
\bibitem [{\citenamefont {Broholm}\ \emph {et~al.}(2020)\citenamefont {Broholm}, \citenamefont {Cava}, \citenamefont {Kivelson}, \citenamefont {Nocera}, \citenamefont {Norman},\ and\ \citenamefont {Senthil}}]{Broholm2020}%
  \BibitemOpen
  \bibfield  {author} {\bibinfo {author} {\bibfnamefont {C.}~\bibnamefont {Broholm}}, \bibinfo {author} {\bibfnamefont {R.~J.}\ \bibnamefont {Cava}}, \bibinfo {author} {\bibfnamefont {S.~A.}\ \bibnamefont {Kivelson}}, \bibinfo {author} {\bibfnamefont {D.~G.}\ \bibnamefont {Nocera}}, \bibinfo {author} {\bibfnamefont {M.~R.}\ \bibnamefont {Norman}},\ and\ \bibinfo {author} {\bibfnamefont {T.}~\bibnamefont {Senthil}},\ }\href {https://science.sciencemag.org/content/367/6475/eaay0668} {\bibfield  {journal} {\bibinfo  {journal} {Science}\ }\textbf {\bibinfo {volume} {367}} (\bibinfo {year} {2020})}\BibitemShut {NoStop}%
\bibitem [{\citenamefont {Takagi}\ \emph {et~al.}(2019)\citenamefont {Takagi}, \citenamefont {Takayama}, \citenamefont {Jackeli}, \citenamefont {Khaliullin},\ and\ \citenamefont {Nagler}}]{Takagi2019}%
  \BibitemOpen
  \bibfield  {author} {\bibinfo {author} {\bibfnamefont {H.}~\bibnamefont {Takagi}}, \bibinfo {author} {\bibfnamefont {T.}~\bibnamefont {Takayama}}, \bibinfo {author} {\bibfnamefont {G.}~\bibnamefont {Jackeli}}, \bibinfo {author} {\bibfnamefont {G.}~\bibnamefont {Khaliullin}},\ and\ \bibinfo {author} {\bibfnamefont {S.~E.}\ \bibnamefont {Nagler}},\ }\href {https://doi.org/10.1038/s42254-019-0038-2} {\bibfield  {journal} {\bibinfo  {journal} {Nature Reviews Physics}\ }\textbf {\bibinfo {volume} {1}},\ \bibinfo {pages} {264} (\bibinfo {year} {2019})}\BibitemShut {NoStop}%
\bibitem [{\citenamefont {Trebst}\ and\ \citenamefont {Hickey}(2022)}]{Trebst2022}%
  \BibitemOpen
  \bibfield  {author} {\bibinfo {author} {\bibfnamefont {S.}~\bibnamefont {Trebst}}\ and\ \bibinfo {author} {\bibfnamefont {C.}~\bibnamefont {Hickey}},\ }\href {https://doi.org/https://doi.org/10.1016/j.physrep.2021.11.003} {\bibfield  {journal} {\bibinfo  {journal} {Physics Reports}\ }\textbf {\bibinfo {volume} {950}},\ \bibinfo {pages} {1} (\bibinfo {year} {2022})}\BibitemShut {NoStop}%
\bibitem [{\citenamefont {Baskaran}\ \emph {et~al.}(2008)\citenamefont {Baskaran}, \citenamefont {Sen},\ and\ \citenamefont {Shankar}}]{Baskaran2008PRB}%
  \BibitemOpen
  \bibfield  {author} {\bibinfo {author} {\bibfnamefont {G.}~\bibnamefont {Baskaran}}, \bibinfo {author} {\bibfnamefont {D.}~\bibnamefont {Sen}},\ and\ \bibinfo {author} {\bibfnamefont {R.}~\bibnamefont {Shankar}},\ }\href {https://doi.org/10.1103/PhysRevB.78.115116} {\bibfield  {journal} {\bibinfo  {journal} {Phys. Rev. B}\ }\textbf {\bibinfo {volume} {78}},\ \bibinfo {pages} {115116} (\bibinfo {year} {2008})}\BibitemShut {NoStop}%
\bibitem [{\citenamefont {Rousochatzakis}\ \emph {et~al.}(2018)\citenamefont {Rousochatzakis}, \citenamefont {Sizyuk},\ and\ \citenamefont {Perkins}}]{Rousochatzakis2018NC}%
  \BibitemOpen
  \bibfield  {author} {\bibinfo {author} {\bibfnamefont {I.}~\bibnamefont {Rousochatzakis}}, \bibinfo {author} {\bibfnamefont {Y.}~\bibnamefont {Sizyuk}},\ and\ \bibinfo {author} {\bibfnamefont {N.~B.}\ \bibnamefont {Perkins}},\ }\href {https://doi.org/10.1038/s41467-018-03934-1} {\bibfield  {journal} {\bibinfo  {journal} {Nat. Commun.}\ }\textbf {\bibinfo {volume} {9}},\ \bibinfo {pages} {1575} (\bibinfo {year} {2018})}\BibitemShut {NoStop}%
\bibitem [{\citenamefont {Jin}\ \emph {et~al.}(2022)\citenamefont {Jin}, \citenamefont {Natori}, \citenamefont {Pollmann},\ and\ \citenamefont {Knolle}}]{Jin2022}%
  \BibitemOpen
  \bibfield  {author} {\bibinfo {author} {\bibfnamefont {H.-K.}\ \bibnamefont {Jin}}, \bibinfo {author} {\bibfnamefont {W.~M.~H.}\ \bibnamefont {Natori}}, \bibinfo {author} {\bibfnamefont {F.}~\bibnamefont {Pollmann}},\ and\ \bibinfo {author} {\bibfnamefont {J.}~\bibnamefont {Knolle}},\ }\href {https://doi.org/10.1038/s41467-022-31503-0} {\bibfield  {journal} {\bibinfo  {journal} {Nature Communications}\ }\textbf {\bibinfo {volume} {13}},\ \bibinfo {pages} {3813} (\bibinfo {year} {2022})}\BibitemShut {NoStop}%
\bibitem [{\citenamefont {Natori}\ \emph {et~al.}(2023)\citenamefont {Natori}, \citenamefont {Jin},\ and\ \citenamefont {Knolle}}]{Natori2023}%
  \BibitemOpen
  \bibfield  {author} {\bibinfo {author} {\bibfnamefont {W.~M.~H.}\ \bibnamefont {Natori}}, \bibinfo {author} {\bibfnamefont {H.-K.}\ \bibnamefont {Jin}},\ and\ \bibinfo {author} {\bibfnamefont {J.}~\bibnamefont {Knolle}},\ }\href {https://doi.org/10.1103/PhysRevB.108.075111} {\bibfield  {journal} {\bibinfo  {journal} {Phys. Rev. B}\ }\textbf {\bibinfo {volume} {108}},\ \bibinfo {pages} {075111} (\bibinfo {year} {2023})}\BibitemShut {NoStop}%
\bibitem [{\citenamefont {de~Carvalho}\ \emph {et~al.}(2023)\citenamefont {de~Carvalho}, \citenamefont {Freire},\ and\ \citenamefont {Pereira}}]{Carvalho2023}%
  \BibitemOpen
  \bibfield  {author} {\bibinfo {author} {\bibfnamefont {V.~S.}\ \bibnamefont {de~Carvalho}}, \bibinfo {author} {\bibfnamefont {H.}~\bibnamefont {Freire}},\ and\ \bibinfo {author} {\bibfnamefont {R.~G.}\ \bibnamefont {Pereira}},\ }\href {https://doi.org/10.1103/PhysRevB.108.094418} {\bibfield  {journal} {\bibinfo  {journal} {Phys. Rev. B}\ }\textbf {\bibinfo {volume} {108}},\ \bibinfo {pages} {094418} (\bibinfo {year} {2023})}\BibitemShut {NoStop}%
\bibitem [{\citenamefont {Ma}(2023)}]{HanMa2023}%
  \BibitemOpen
  \bibfield  {author} {\bibinfo {author} {\bibfnamefont {H.}~\bibnamefont {Ma}},\ }\href {https://doi.org/10.1103/PhysRevLett.130.156701} {\bibfield  {journal} {\bibinfo  {journal} {Phys. Rev. Lett.}\ }\textbf {\bibinfo {volume} {130}},\ \bibinfo {pages} {156701} (\bibinfo {year} {2023})}\BibitemShut {NoStop}%
\bibitem [{\citenamefont {Georgiou}\ \emph {et~al.}(2024)\citenamefont {Georgiou}, \citenamefont {Rousochatzakis}, \citenamefont {Farnell}, \citenamefont {Richter},\ and\ \citenamefont {Bishop}}]{Georgiou2024}%
  \BibitemOpen
  \bibfield  {author} {\bibinfo {author} {\bibfnamefont {M.}~\bibnamefont {Georgiou}}, \bibinfo {author} {\bibfnamefont {I.}~\bibnamefont {Rousochatzakis}}, \bibinfo {author} {\bibfnamefont {D.~J.~J.}\ \bibnamefont {Farnell}}, \bibinfo {author} {\bibfnamefont {J.}~\bibnamefont {Richter}},\ and\ \bibinfo {author} {\bibfnamefont {R.~F.}\ \bibnamefont {Bishop}},\ }\href {https://arxiv.org/abs/2405.14378} {\bibfield  {journal} {\bibinfo  {journal} {arXiv:2405.14378}\ } (\bibinfo {year} {2024})}\BibitemShut {NoStop}%
\bibitem [{\citenamefont {Hermanns}\ \emph {et~al.}(2018)\citenamefont {Hermanns}, \citenamefont {Kimchi},\ and\ \citenamefont {Knolle}}]{Knolle2017}%
  \BibitemOpen
  \bibfield  {author} {\bibinfo {author} {\bibfnamefont {M.}~\bibnamefont {Hermanns}}, \bibinfo {author} {\bibfnamefont {I.}~\bibnamefont {Kimchi}},\ and\ \bibinfo {author} {\bibfnamefont {J.}~\bibnamefont {Knolle}},\ }\href {https://doi.org/10.1146/annurev-conmatphys-033117-053934} {\bibfield  {journal} {\bibinfo  {journal} {Annual Review of Condensed Matter Physics}\ }\textbf {\bibinfo {volume} {9}},\ \bibinfo {pages} {17} (\bibinfo {year} {2018})}\BibitemShut {NoStop}%
\bibitem [{\citenamefont {Rousochatzakis}\ \emph {et~al.}(2024)\citenamefont {Rousochatzakis}, \citenamefont {Perkins}, \citenamefont {Luo},\ and\ \citenamefont {Kee}}]{Rousochatzakis2024}%
  \BibitemOpen
  \bibfield  {author} {\bibinfo {author} {\bibfnamefont {I.}~\bibnamefont {Rousochatzakis}}, \bibinfo {author} {\bibfnamefont {N.~B.}\ \bibnamefont {Perkins}}, \bibinfo {author} {\bibfnamefont {Q.}~\bibnamefont {Luo}},\ and\ \bibinfo {author} {\bibfnamefont {H.-Y.}\ \bibnamefont {Kee}},\ }\href {https://doi.org/10.1088/1361-6633/ad208d} {\bibfield  {journal} {\bibinfo  {journal} {Reports on Progress in Physics}\ }\textbf {\bibinfo {volume} {87}},\ \bibinfo {pages} {026502} (\bibinfo {year} {2024})}\BibitemShut {NoStop}%
\bibitem [{\citenamefont {Jackeli}\ and\ \citenamefont {Khaliullin}(2009)}]{Jackeli2009}%
  \BibitemOpen
  \bibfield  {author} {\bibinfo {author} {\bibfnamefont {G.}~\bibnamefont {Jackeli}}\ and\ \bibinfo {author} {\bibfnamefont {G.}~\bibnamefont {Khaliullin}},\ }\href {https://doi.org/10.1103/PhysRevLett.102.017205} {\bibfield  {journal} {\bibinfo  {journal} {Physical Review Letters}\ }\textbf {\bibinfo {volume} {102}},\ \bibinfo {pages} {017205} (\bibinfo {year} {2009})}\BibitemShut {NoStop}%
\bibitem [{\citenamefont {Chaloupka}\ \emph {et~al.}(2010)\citenamefont {Chaloupka}, \citenamefont {Jackeli},\ and\ \citenamefont {Khaliullin}}]{Chaloupka2010}%
  \BibitemOpen
  \bibfield  {author} {\bibinfo {author} {\bibfnamefont {J.}~\bibnamefont {Chaloupka}}, \bibinfo {author} {\bibfnamefont {G.}~\bibnamefont {Jackeli}},\ and\ \bibinfo {author} {\bibfnamefont {G.}~\bibnamefont {Khaliullin}},\ }\href {https://doi.org/10.1103/PhysRevLett.105.027204} {\bibfield  {journal} {\bibinfo  {journal} {Physical Review Letters}\ }\textbf {\bibinfo {volume} {105}},\ \bibinfo {pages} {027204} (\bibinfo {year} {2010})}\BibitemShut {NoStop}%
\bibitem [{\citenamefont {Khaliullin}(2005)}]{Khaliullin2005}%
  \BibitemOpen
  \bibfield  {author} {\bibinfo {author} {\bibfnamefont {G.}~\bibnamefont {Khaliullin}},\ }\href {https://doi.org/10.1143/PTPS.160.155} {\bibfield  {journal} {\bibinfo  {journal} {Progr. Theor. Phys. Suppl.}\ }\textbf {\bibinfo {volume} {160}},\ \bibinfo {pages} {155} (\bibinfo {year} {2005})}\BibitemShut {NoStop}%
\bibitem [{\citenamefont {Plumb}\ \emph {et~al.}(2014)\citenamefont {Plumb}, \citenamefont {Clancy}, \citenamefont {Sandilands}, \citenamefont {Shankar}, \citenamefont {Hu}, \citenamefont {Burch}, \citenamefont {Kee},\ and\ \citenamefont {Kim}}]{Plumb2014}%
  \BibitemOpen
  \bibfield  {author} {\bibinfo {author} {\bibfnamefont {K.~W.}\ \bibnamefont {Plumb}}, \bibinfo {author} {\bibfnamefont {J.~P.}\ \bibnamefont {Clancy}}, \bibinfo {author} {\bibfnamefont {L.~J.}\ \bibnamefont {Sandilands}}, \bibinfo {author} {\bibfnamefont {V.~V.}\ \bibnamefont {Shankar}}, \bibinfo {author} {\bibfnamefont {Y.~F.}\ \bibnamefont {Hu}}, \bibinfo {author} {\bibfnamefont {K.~S.}\ \bibnamefont {Burch}}, \bibinfo {author} {\bibfnamefont {H.-Y.}\ \bibnamefont {Kee}},\ and\ \bibinfo {author} {\bibfnamefont {Y.-J.}\ \bibnamefont {Kim}},\ }\href {https://doi.org/10.1103/PhysRevB.90.041112} {\bibfield  {journal} {\bibinfo  {journal} {Phys. Rev. B}\ }\textbf {\bibinfo {volume} {90}},\ \bibinfo {pages} {041112(R)} (\bibinfo {year} {2014})}\BibitemShut {NoStop}%
\bibitem [{\citenamefont {Banerjee}\ \emph {et~al.}(2017)\citenamefont {Banerjee}, \citenamefont {Yan}, \citenamefont {Knolle}, \citenamefont {Bridges}, \citenamefont {Stone}, \citenamefont {Lumsden}, \citenamefont {Mandrus}, \citenamefont {Tennant}, \citenamefont {Moessner},\ and\ \citenamefont {Nagler}}]{banerjee_neutron_2017}%
  \BibitemOpen
  \bibfield  {author} {\bibinfo {author} {\bibfnamefont {A.}~\bibnamefont {Banerjee}}, \bibinfo {author} {\bibfnamefont {J.}~\bibnamefont {Yan}}, \bibinfo {author} {\bibfnamefont {J.}~\bibnamefont {Knolle}}, \bibinfo {author} {\bibfnamefont {C.~A.}\ \bibnamefont {Bridges}}, \bibinfo {author} {\bibfnamefont {M.~B.}\ \bibnamefont {Stone}}, \bibinfo {author} {\bibfnamefont {M.~D.}\ \bibnamefont {Lumsden}}, \bibinfo {author} {\bibfnamefont {D.~G.}\ \bibnamefont {Mandrus}}, \bibinfo {author} {\bibfnamefont {D.~A.}\ \bibnamefont {Tennant}}, \bibinfo {author} {\bibfnamefont {R.}~\bibnamefont {Moessner}},\ and\ \bibinfo {author} {\bibfnamefont {S.~E.}\ \bibnamefont {Nagler}},\ }\href {https://doi.org/10.1126/science.aah6015} {\bibfield  {journal} {\bibinfo  {journal} {Science}\ }\textbf {\bibinfo {volume} {356}},\ \bibinfo {pages} {1055} (\bibinfo {year} {2017})}\BibitemShut {NoStop}%
\bibitem [{\citenamefont {Do}\ \emph {et~al.}(2017)\citenamefont {Do}, \citenamefont {Park}, \citenamefont {Yoshitake}, \citenamefont {Nasu}, \citenamefont {Motome}, \citenamefont {Kwon}, \citenamefont {Adroja}, \citenamefont {Voneshen}, \citenamefont {Kim}, \citenamefont {Jang}, \citenamefont {Park}, \citenamefont {Choi},\ and\ \citenamefont {Ji}}]{do_majorana_2017}%
  \BibitemOpen
  \bibfield  {author} {\bibinfo {author} {\bibfnamefont {S.-H.}\ \bibnamefont {Do}}, \bibinfo {author} {\bibfnamefont {S.-Y.}\ \bibnamefont {Park}}, \bibinfo {author} {\bibfnamefont {J.}~\bibnamefont {Yoshitake}}, \bibinfo {author} {\bibfnamefont {J.}~\bibnamefont {Nasu}}, \bibinfo {author} {\bibfnamefont {Y.}~\bibnamefont {Motome}}, \bibinfo {author} {\bibfnamefont {Y.}~\bibnamefont {Kwon}}, \bibinfo {author} {\bibfnamefont {D.~T.}\ \bibnamefont {Adroja}}, \bibinfo {author} {\bibfnamefont {D.~J.}\ \bibnamefont {Voneshen}}, \bibinfo {author} {\bibfnamefont {K.}~\bibnamefont {Kim}}, \bibinfo {author} {\bibfnamefont {T.-H.}\ \bibnamefont {Jang}}, \bibinfo {author} {\bibfnamefont {J.-H.}\ \bibnamefont {Park}}, \bibinfo {author} {\bibfnamefont {K.-Y.}\ \bibnamefont {Choi}},\ and\ \bibinfo {author} {\bibfnamefont {S.}~\bibnamefont {Ji}},\ }\href {https://doi.org/10.1038/nphys4264} {\bibfield  {journal} {\bibinfo  {journal} {Nat. Phys.}\ }\textbf {\bibinfo {volume} {13}},\ \bibinfo {pages} {1079} (\bibinfo {year}
  {2017})}\BibitemShut {NoStop}%
\bibitem [{\citenamefont {Janša}\ \emph {et~al.}(2018)\citenamefont {Janša}, \citenamefont {Zorko}, \citenamefont {Gomilšek}, \citenamefont {Pregelj}, \citenamefont {Krämer}, \citenamefont {Biner}, \citenamefont {Biffin}, \citenamefont {Rüegg},\ and\ \citenamefont {Klanjšek}}]{jansa_observation_2018}%
  \BibitemOpen
  \bibfield  {author} {\bibinfo {author} {\bibfnamefont {N.}~\bibnamefont {Janša}}, \bibinfo {author} {\bibfnamefont {A.}~\bibnamefont {Zorko}}, \bibinfo {author} {\bibfnamefont {M.}~\bibnamefont {Gomilšek}}, \bibinfo {author} {\bibfnamefont {M.}~\bibnamefont {Pregelj}}, \bibinfo {author} {\bibfnamefont {K.~W.}\ \bibnamefont {Krämer}}, \bibinfo {author} {\bibfnamefont {D.}~\bibnamefont {Biner}}, \bibinfo {author} {\bibfnamefont {A.}~\bibnamefont {Biffin}}, \bibinfo {author} {\bibfnamefont {C.}~\bibnamefont {Rüegg}},\ and\ \bibinfo {author} {\bibfnamefont {M.}~\bibnamefont {Klanjšek}},\ }\href {https://doi.org/10.1038/s41567-018-0129-5} {\bibfield  {journal} {\bibinfo  {journal} {Nat. Phys.}\ }\textbf {\bibinfo {volume} {14}},\ \bibinfo {pages} {786} (\bibinfo {year} {2018})}\BibitemShut {NoStop}%
\bibitem [{\citenamefont {Xu}\ \emph {et~al.}(2020)\citenamefont {Xu}, \citenamefont {Feng}, \citenamefont {Kawamura}, \citenamefont {Yamaji}, \citenamefont {Nahas}, \citenamefont {Prokhorenko}, \citenamefont {Qi}, \citenamefont {Xiang},\ and\ \citenamefont {Bellaiche}}]{Xu2020}%
  \BibitemOpen
  \bibfield  {author} {\bibinfo {author} {\bibfnamefont {C.}~\bibnamefont {Xu}}, \bibinfo {author} {\bibfnamefont {J.}~\bibnamefont {Feng}}, \bibinfo {author} {\bibfnamefont {M.}~\bibnamefont {Kawamura}}, \bibinfo {author} {\bibfnamefont {Y.}~\bibnamefont {Yamaji}}, \bibinfo {author} {\bibfnamefont {Y.}~\bibnamefont {Nahas}}, \bibinfo {author} {\bibfnamefont {S.}~\bibnamefont {Prokhorenko}}, \bibinfo {author} {\bibfnamefont {Y.}~\bibnamefont {Qi}}, \bibinfo {author} {\bibfnamefont {H.}~\bibnamefont {Xiang}},\ and\ \bibinfo {author} {\bibfnamefont {L.}~\bibnamefont {Bellaiche}},\ }\href {https://doi.org/10.1103/PhysRevLett.124.087205} {\bibfield  {journal} {\bibinfo  {journal} {Phys. Rev. Lett.}\ }\textbf {\bibinfo {volume} {124}},\ \bibinfo {pages} {087205} (\bibinfo {year} {2020})}\BibitemShut {NoStop}%
\bibitem [{\citenamefont {Lee}\ \emph {et~al.}(2020)\citenamefont {Lee}, \citenamefont {Utermohlen}, \citenamefont {Weber}, \citenamefont {Hwang}, \citenamefont {Zhang}, \citenamefont {van Tol}, \citenamefont {Goldberger}, \citenamefont {Trivedi},\ and\ \citenamefont {Hammel}}]{Lee2020}%
  \BibitemOpen
  \bibfield  {author} {\bibinfo {author} {\bibfnamefont {I.}~\bibnamefont {Lee}}, \bibinfo {author} {\bibfnamefont {F.~G.}\ \bibnamefont {Utermohlen}}, \bibinfo {author} {\bibfnamefont {D.}~\bibnamefont {Weber}}, \bibinfo {author} {\bibfnamefont {K.}~\bibnamefont {Hwang}}, \bibinfo {author} {\bibfnamefont {C.}~\bibnamefont {Zhang}}, \bibinfo {author} {\bibfnamefont {J.}~\bibnamefont {van Tol}}, \bibinfo {author} {\bibfnamefont {J.~E.}\ \bibnamefont {Goldberger}}, \bibinfo {author} {\bibfnamefont {N.}~\bibnamefont {Trivedi}},\ and\ \bibinfo {author} {\bibfnamefont {P.~C.}\ \bibnamefont {Hammel}},\ }\href {https://doi.org/10.1103/PhysRevLett.124.017201} {\bibfield  {journal} {\bibinfo  {journal} {Phys. Rev. Lett.}\ }\textbf {\bibinfo {volume} {124}},\ \bibinfo {pages} {017201} (\bibinfo {year} {2020})}\BibitemShut {NoStop}%
\bibitem [{\citenamefont {Stavropoulos}\ \emph {et~al.}(2019)\citenamefont {Stavropoulos}, \citenamefont {Pereira},\ and\ \citenamefont {Kee}}]{Stavropoulos2019}%
  \BibitemOpen
  \bibfield  {author} {\bibinfo {author} {\bibfnamefont {P.~P.}\ \bibnamefont {Stavropoulos}}, \bibinfo {author} {\bibfnamefont {D.}~\bibnamefont {Pereira}},\ and\ \bibinfo {author} {\bibfnamefont {H.-Y.}\ \bibnamefont {Kee}},\ }\href {https://doi.org/10.1103/PhysRevLett.123.037203} {\bibfield  {journal} {\bibinfo  {journal} {Phys. Rev. Lett.}\ }\textbf {\bibinfo {volume} {123}},\ \bibinfo {pages} {037203} (\bibinfo {year} {2019})}\BibitemShut {NoStop}%
\bibitem [{\citenamefont {Stavropoulos}\ \emph {et~al.}(2021)\citenamefont {Stavropoulos}, \citenamefont {Liu},\ and\ \citenamefont {Kee}}]{Stavropoulos2021}%
  \BibitemOpen
  \bibfield  {author} {\bibinfo {author} {\bibfnamefont {P.~P.}\ \bibnamefont {Stavropoulos}}, \bibinfo {author} {\bibfnamefont {X.}~\bibnamefont {Liu}},\ and\ \bibinfo {author} {\bibfnamefont {H.-Y.}\ \bibnamefont {Kee}},\ }\href {https://doi.org/10.1103/PhysRevResearch.3.013216} {\bibfield  {journal} {\bibinfo  {journal} {Phys. Rev. Res.}\ }\textbf {\bibinfo {volume} {3}},\ \bibinfo {pages} {013216} (\bibinfo {year} {2021})}\BibitemShut {NoStop}%
\bibitem [{\citenamefont {Yamada}\ \emph {et~al.}(2018)\citenamefont {Yamada}, \citenamefont {Oshikawa},\ and\ \citenamefont {Jackeli}}]{Yamada2018}%
  \BibitemOpen
  \bibfield  {author} {\bibinfo {author} {\bibfnamefont {M.~G.}\ \bibnamefont {Yamada}}, \bibinfo {author} {\bibfnamefont {M.}~\bibnamefont {Oshikawa}},\ and\ \bibinfo {author} {\bibfnamefont {G.}~\bibnamefont {Jackeli}},\ }\href {https://doi.org/10.1103/PhysRevLett.121.097201} {\bibfield  {journal} {\bibinfo  {journal} {Phys. Rev. Lett.}\ }\textbf {\bibinfo {volume} {121}},\ \bibinfo {pages} {097201} (\bibinfo {year} {2018})}\BibitemShut {NoStop}%
\bibitem [{\citenamefont {Yamada}\ \emph {et~al.}(2021)\citenamefont {Yamada}, \citenamefont {Oshikawa},\ and\ \citenamefont {Jackeli}}]{Yamada2021}%
  \BibitemOpen
  \bibfield  {author} {\bibinfo {author} {\bibfnamefont {M.~G.}\ \bibnamefont {Yamada}}, \bibinfo {author} {\bibfnamefont {M.}~\bibnamefont {Oshikawa}},\ and\ \bibinfo {author} {\bibfnamefont {G.}~\bibnamefont {Jackeli}},\ }\href {https://doi.org/10.1103/PhysRevB.104.224436} {\bibfield  {journal} {\bibinfo  {journal} {Phys. Rev. B}\ }\textbf {\bibinfo {volume} {104}},\ \bibinfo {pages} {224436} (\bibinfo {year} {2021})}\BibitemShut {NoStop}%
\bibitem [{\citenamefont {Natori}\ \emph {et~al.}(2018)\citenamefont {Natori}, \citenamefont {Andrade},\ and\ \citenamefont {Pereira}}]{Natori2018}%
  \BibitemOpen
  \bibfield  {author} {\bibinfo {author} {\bibfnamefont {W.~M.~H.}\ \bibnamefont {Natori}}, \bibinfo {author} {\bibfnamefont {E.~C.}\ \bibnamefont {Andrade}},\ and\ \bibinfo {author} {\bibfnamefont {R.~G.}\ \bibnamefont {Pereira}},\ }\href {https://doi.org/10.1103/PhysRevB.98.195113} {\bibfield  {journal} {\bibinfo  {journal} {Phys. Rev. B}\ }\textbf {\bibinfo {volume} {98}},\ \bibinfo {pages} {195113} (\bibinfo {year} {2018})}\BibitemShut {NoStop}%
\bibitem [{\citenamefont {Churchill}\ \emph {et~al.}(2024)\citenamefont {Churchill}, \citenamefont {Zhang},\ and\ \citenamefont {Kee}}]{Churchill_arXiv2024}%
  \BibitemOpen
  \bibfield  {author} {\bibinfo {author} {\bibfnamefont {D.}~\bibnamefont {Churchill}}, \bibinfo {author} {\bibfnamefont {E.~Z.}\ \bibnamefont {Zhang}},\ and\ \bibinfo {author} {\bibfnamefont {H.-Y.}\ \bibnamefont {Kee}},\ }\href {https://arxiv.org/abs/2410.21389} {\bibinfo {title} {Microscopic roadmap to a yao-lee spin-orbital liquid}} (\bibinfo {year} {2024}),\ \Eprint {https://arxiv.org/abs/2410.21389} {arXiv:2410.21389 [cond-mat.str-el]} \BibitemShut {NoStop}%
\bibitem [{\citenamefont {Swaroop}\ and\ \citenamefont {Flengas}(1964{\natexlab{a}})}]{Swaroop1964}%
  \BibitemOpen
  \bibfield  {author} {\bibinfo {author} {\bibfnamefont {B.}~\bibnamefont {Swaroop}}\ and\ \bibinfo {author} {\bibfnamefont {S.~N.}\ \bibnamefont {Flengas}},\ }\href {https://doi.org/10.1139/v64-228} {\bibfield  {journal} {\bibinfo  {journal} {Canadian Journal of Chemistry}\ }\textbf {\bibinfo {volume} {42}},\ \bibinfo {pages} {1495} (\bibinfo {year} {1964}{\natexlab{a}})}\BibitemShut {NoStop}%
\bibitem [{\citenamefont {Swaroop}\ and\ \citenamefont {Flengas}(1964{\natexlab{b}})}]{Swaroop1964-2}%
  \BibitemOpen
  \bibfield  {author} {\bibinfo {author} {\bibfnamefont {B.}~\bibnamefont {Swaroop}}\ and\ \bibinfo {author} {\bibfnamefont {S.~N.}\ \bibnamefont {Flengas}},\ }\href {https://doi.org/10.1139/p64-177} {\bibfield  {journal} {\bibinfo  {journal} {Canadian Journal of Chemistry}\ }\textbf {\bibinfo {volume} {42}},\ \bibinfo {pages} {1886} (\bibinfo {year} {1964}{\natexlab{b}})}\BibitemShut {NoStop}%
\bibitem [{\citenamefont {Chen}\ \emph {et~al.}(2010)\citenamefont {Chen}, \citenamefont {Pereira},\ and\ \citenamefont {Balents}}]{Chen2010}%
  \BibitemOpen
  \bibfield  {author} {\bibinfo {author} {\bibfnamefont {G.}~\bibnamefont {Chen}}, \bibinfo {author} {\bibfnamefont {R.}~\bibnamefont {Pereira}},\ and\ \bibinfo {author} {\bibfnamefont {L.}~\bibnamefont {Balents}},\ }\href {https://doi.org/10.1103/PhysRevB.82.174440} {\bibfield  {journal} {\bibinfo  {journal} {Phys. Rev. B}\ }\textbf {\bibinfo {volume} {82}},\ \bibinfo {pages} {174440} (\bibinfo {year} {2010})}\BibitemShut {NoStop}%
\bibitem [{\citenamefont {Néel}(1952)}]{Néel_1952}%
  \BibitemOpen
  \bibfield  {author} {\bibinfo {author} {\bibfnamefont {L.}~\bibnamefont {Néel}},\ }\href {https://doi.org/10.1088/0370-1298/65/11/301} {\bibfield  {journal} {\bibinfo  {journal} {Proceedings of the Physical Society. Section A}\ }\textbf {\bibinfo {volume} {65}},\ \bibinfo {pages} {869} (\bibinfo {year} {1952})}\BibitemShut {NoStop}%
\bibitem [{\citenamefont {Wolf}(1961)}]{Wolf_1961}%
  \BibitemOpen
  \bibfield  {author} {\bibinfo {author} {\bibfnamefont {W.~P.}\ \bibnamefont {Wolf}},\ }\href {https://doi.org/10.1088/0034-4885/24/1/306} {\bibfield  {journal} {\bibinfo  {journal} {Reports on Progress in Physics}\ }\textbf {\bibinfo {volume} {24}},\ \bibinfo {pages} {212} (\bibinfo {year} {1961})}\BibitemShut {NoStop}%
\bibitem [{\citenamefont {Takahashi}\ \emph {et~al.}(2024)\citenamefont {Takahashi}, \citenamefont {Kao}, \citenamefont {Fujimoto},\ and\ \citenamefont {Perkins}}]{Masahiro2024}%
  \BibitemOpen
  \bibfield  {author} {\bibinfo {author} {\bibfnamefont {M.~O.}\ \bibnamefont {Takahashi}}, \bibinfo {author} {\bibfnamefont {W.-H.}\ \bibnamefont {Kao}}, \bibinfo {author} {\bibfnamefont {S.}~\bibnamefont {Fujimoto}},\ and\ \bibinfo {author} {\bibfnamefont {N.~B.}\ \bibnamefont {Perkins}},\ }\href {https://doi.org/10.48550/arXiv.2409.02190} {\bibfield  {journal} {\bibinfo  {journal} {arXiv:2409.02190}\ } (\bibinfo {year} {2024})}\BibitemShut {NoStop}%
\bibitem [{\citenamefont {Wang}\ and\ \citenamefont {Vishwanath}(2009)}]{Wang2009}%
  \BibitemOpen
  \bibfield  {author} {\bibinfo {author} {\bibfnamefont {F.}~\bibnamefont {Wang}}\ and\ \bibinfo {author} {\bibfnamefont {A.}~\bibnamefont {Vishwanath}},\ }\href {https://doi.org/10.1103/PhysRevB.80.064413} {\bibfield  {journal} {\bibinfo  {journal} {Phys. Rev. B}\ }\textbf {\bibinfo {volume} {80}},\ \bibinfo {pages} {064413} (\bibinfo {year} {2009})}\BibitemShut {NoStop}%
\bibitem [{\citenamefont {Coleman}\ \emph {et~al.}(1994)\citenamefont {Coleman}, \citenamefont {Miranda},\ and\ \citenamefont {Tsvelik}}]{Coleman1994}%
  \BibitemOpen
  \bibfield  {author} {\bibinfo {author} {\bibfnamefont {P.}~\bibnamefont {Coleman}}, \bibinfo {author} {\bibfnamefont {E.}~\bibnamefont {Miranda}},\ and\ \bibinfo {author} {\bibfnamefont {A.}~\bibnamefont {Tsvelik}},\ }\href {https://doi.org/10.1103/PhysRevB.49.8955} {\bibfield  {journal} {\bibinfo  {journal} {Phys. Rev. B}\ }\textbf {\bibinfo {volume} {49}},\ \bibinfo {pages} {8955} (\bibinfo {year} {1994})}\BibitemShut {NoStop}%
\bibitem [{\citenamefont {Fu}\ \emph {et~al.}(2018)\citenamefont {Fu}, \citenamefont {Knolle},\ and\ \citenamefont {Perkins}}]{Fu2018}%
  \BibitemOpen
  \bibfield  {author} {\bibinfo {author} {\bibfnamefont {J.}~\bibnamefont {Fu}}, \bibinfo {author} {\bibfnamefont {J.}~\bibnamefont {Knolle}},\ and\ \bibinfo {author} {\bibfnamefont {N.~B.}\ \bibnamefont {Perkins}},\ }\href {https://doi.org/10.1103/PhysRevB.97.115142} {\bibfield  {journal} {\bibinfo  {journal} {Phys. Rev. B}\ }\textbf {\bibinfo {volume} {97}},\ \bibinfo {pages} {115142} (\bibinfo {year} {2018})}\BibitemShut {NoStop}%
\bibitem [{\citenamefont {Schaden}\ and\ \citenamefont {Reuther}(2023)}]{Schaden2023}%
  \BibitemOpen
  \bibfield  {author} {\bibinfo {author} {\bibfnamefont {Y.}~\bibnamefont {Schaden}}\ and\ \bibinfo {author} {\bibfnamefont {J.}~\bibnamefont {Reuther}},\ }\href {https://doi.org/10.1103/PhysRevResearch.5.023067} {\bibfield  {journal} {\bibinfo  {journal} {Phys. Rev. Res.}\ }\textbf {\bibinfo {volume} {5}},\ \bibinfo {pages} {023067} (\bibinfo {year} {2023})}\BibitemShut {NoStop}%
\bibitem [{\citenamefont {Lieb}(1994)}]{Lieb1994}%
  \BibitemOpen
  \bibfield  {author} {\bibinfo {author} {\bibfnamefont {E.~H.}\ \bibnamefont {Lieb}},\ }\href {https://doi.org/10.1103/PhysRevLett.73.2158} {\bibfield  {journal} {\bibinfo  {journal} {Phys. Rev. Lett.}\ }\textbf {\bibinfo {volume} {73}},\ \bibinfo {pages} {2158} (\bibinfo {year} {1994})}\BibitemShut {NoStop}%
\bibitem [{\citenamefont {Ralko}\ and\ \citenamefont {Merino}(2020)}]{Ralko2020}%
  \BibitemOpen
  \bibfield  {author} {\bibinfo {author} {\bibfnamefont {A.}~\bibnamefont {Ralko}}\ and\ \bibinfo {author} {\bibfnamefont {J.}~\bibnamefont {Merino}},\ }\href {https://doi.org/10.1103/PhysRevLett.124.217203} {\bibfield  {journal} {\bibinfo  {journal} {Phys. Rev. Lett.}\ }\textbf {\bibinfo {volume} {124}},\ \bibinfo {pages} {217203} (\bibinfo {year} {2020})}\BibitemShut {NoStop}%
\bibitem [{\citenamefont {White}(1992)}]{White1992}%
  \BibitemOpen
  \bibfield  {author} {\bibinfo {author} {\bibfnamefont {S.~R.}\ \bibnamefont {White}},\ }\href@noop {} {\bibfield  {journal} {\bibinfo  {journal} {Physical review letters}\ }\textbf {\bibinfo {volume} {69}},\ \bibinfo {pages} {2863} (\bibinfo {year} {1992})}\BibitemShut {NoStop}%
\bibitem [{\citenamefont {White}(1993)}]{White1993}%
  \BibitemOpen
  \bibfield  {author} {\bibinfo {author} {\bibfnamefont {S.~R.}\ \bibnamefont {White}},\ }\href@noop {} {\bibfield  {journal} {\bibinfo  {journal} {Physical Review B}\ }\textbf {\bibinfo {volume} {48}},\ \bibinfo {pages} {10345} (\bibinfo {year} {1993})}\BibitemShut {NoStop}%
\bibitem [{\citenamefont {Weinberg}\ and\ \citenamefont {Bukov}(2017)}]{quspin}%
  \BibitemOpen
  \bibfield  {author} {\bibinfo {author} {\bibfnamefont {P.}~\bibnamefont {Weinberg}}\ and\ \bibinfo {author} {\bibfnamefont {M.}~\bibnamefont {Bukov}},\ }\href {https://doi.org/10.21468/SciPostPhys.2.1.003} {\bibfield  {journal} {\bibinfo  {journal} {SciPost Phys.}\ }\textbf {\bibinfo {volume} {2}},\ \bibinfo {pages} {003} (\bibinfo {year} {2017})}\BibitemShut {NoStop}%
\bibitem [{\citenamefont {Winter}\ \emph {et~al.}(2016)\citenamefont {Winter}, \citenamefont {Li}, \citenamefont {Jeschke},\ and\ \citenamefont {Valent\'{\i}}}]{Winter2016PRB}%
  \BibitemOpen
  \bibfield  {author} {\bibinfo {author} {\bibfnamefont {S.~M.}\ \bibnamefont {Winter}}, \bibinfo {author} {\bibfnamefont {Y.}~\bibnamefont {Li}}, \bibinfo {author} {\bibfnamefont {H.~O.}\ \bibnamefont {Jeschke}},\ and\ \bibinfo {author} {\bibfnamefont {R.}~\bibnamefont {Valent\'{\i}}},\ }\href {https://doi.org/10.1103/PhysRevB.93.214431} {\bibfield  {journal} {\bibinfo  {journal} {Phys. Rev. B}\ }\textbf {\bibinfo {volume} {93}},\ \bibinfo {pages} {214431} (\bibinfo {year} {2016})}\BibitemShut {NoStop}%
\bibitem [{\citenamefont {Rau}\ \emph {et~al.}(2014)\citenamefont {Rau}, \citenamefont {Lee},\ and\ \citenamefont {Kee}}]{Rau2014}%
  \BibitemOpen
  \bibfield  {author} {\bibinfo {author} {\bibfnamefont {J.~G.}\ \bibnamefont {Rau}}, \bibinfo {author} {\bibfnamefont {E.~K.-H.}\ \bibnamefont {Lee}},\ and\ \bibinfo {author} {\bibfnamefont {H.-Y.}\ \bibnamefont {Kee}},\ }\href {https://doi.org/10.1103/PhysRevLett.112.077204} {\bibfield  {journal} {\bibinfo  {journal} {Physical Review Letters}\ }\textbf {\bibinfo {volume} {112}},\ \bibinfo {pages} {077204} (\bibinfo {year} {2014})}\BibitemShut {NoStop}%
\bibitem [{Note1()}]{Note1}%
  \BibitemOpen
  \bibinfo {note} {The octupolar term $(J^x)^3$ in the Hamiltonian is derived from the orthogonal basis component $(J^x)^3-\protect \frac {41}{20}J_x$, and the remaining dipolar part $-\protect \frac {41}{20}J_x$ is absorbed in the dipolar-dipolar interaction, i,e., $J_x S_x$, $J_x S_y$, and $J_x S_z$.}\BibitemShut {Stop}%
\end{thebibliography}%

\end{document}